%% file: ms.tex
\documentclass[]{emulateapj}

\newcommand{\lya}{Ly-$\alpha$~}

\newcommand{\nciv}{N_{\mciv}}
\newcommand{\pcmsq}{cm$^{-2}$}
\newcommand{\cii}{C~{\sc ii}~}

\newcommand{\civ}{C~{\sc iv}~}
\newcommand{\cv}{C~{\sc v}~}

\newcommand{\ovi}{O~{\sc vi}~}

\newcommand{\siiv}{Si~{\sc iv}~}
\newcommand{\hi}{H~{\sc i}~}

\newcommand{\mgii}{Mg~{\sc ii}~}
\newcommand{\oden}{\rho/\bar{\rho}}

\newcommand{\mciv}{{\rm C \; \mbox{\tiny IV}}}

\newcommand{\kms}{km s$^{-1}$}
\newcommand{\ociv}{\Omega_{\mciv}}

\begin{document}
  
\bibliographystyle{/home/simcoe/latex/apj}
  
\title{High Redshift Intergalactic \civ Abundance Measurements \\ From
The Near-Infrared Spectra of Two $z\sim 6$ QSOs \altaffilmark{1}}
  
\author{Robert A. Simcoe\altaffilmark{2,3}}  

\altaffiltext{1}{ Based on observations obtained at the Gemini
Observatory, which is operated by the Association of Universities for
Research in Astronomy, Inc., under a cooperative agreement with the
NSF on behalf of the Gemini partnership: the National Science
Foundation (United States), the Particle Physics and Astronomy
Research Council (United Kingdom), the National Research Council
(Canada), CONICYT (Chile), the Australian Research Council
(Australia), CNPq (Brazil) and CONICET (Argentina)}
\altaffiltext{2}{MIT Center for Space Research, 77 Massachusetts Ave.
\#37-673, Cambridge, MA 02139, USA; simcoe@mit.edu}
\altaffiltext{3}{Pappalardo Fellow in Physics}

\begin{abstract}

New measurements of the $z\sim 6$ intergalactic \civ abundance are
presented, using moderate resolution IR spectra of two QSOs taken with
GNIRS on Gemini South.  These data were systematically searched for
high redshift \civ absorption lines, using objective selection
criteria.  Comprehensive tests were performed to quantify sample
incompleteness, as well as the rate of false positive \civ
identifications.  The trend of constant $\ociv(z)$ observed at $z\sim
2-5$ appears to continue to $z\sim 6$, the highest observed redshift.
The \civ sample is also consistent with the redshift-invariant form of
the \civ column density distribution reported by
\citet{songaila_omegaz} at lower redshift, although with fairly large
uncertainties due to a smaller sample size and noisier infrared data.
The constant value of $\ociv$ does not necessarily imply that the IGM
was infused with an early metallicity ``floor,'' but the presence of
early \civ does indicate that heavy-element enrichment began $\lesssim
1$ Gyr after the Big Bang.  The lack of a decline in $\ociv$ at high
redshift may indicate that integrated \civ measurements are sensitive
to the instantaneous rate of feedback from galaxy formation at each
epoch.  Alternatively, it could result from a balance in the evolution
of the intergalactic gas density, ionization conditions, and
heavy-element abundance over time.

\end{abstract}

\section{Introduction}\label{sec:introduction}

Recent CMB measurements of the electron scattering optical depth
suggest that star formation first began to ionize the IGM at $z\approx
11$, about $425$ Myr after the Big Bang \citep{spergel_wmap_3year}.
The subsequent epoch between $z=11$ and the end of reionization at
$z\sim 6$
\citep{fan_reionization,white_reionization,george_reionization}
encompasses only $\sim 500$ Myr.  This interval begins to approach the
dynamical timescale for massive galaxy formation and supernova
feedback, so the galaxies formed at $z\sim 6-10$ may be among the
first objects to produce the heavy elements seen in the IGM at $z\sim
3$ \citep{schaye_civ_pixels,simcoe2004}.
 
If the large scale enrichment of the IGM actually began within this
short window, one would expect to see either a significant downward
trend in heavy-element abundances, or at least an increase in
abundance scatter approaching $z\sim 5-6$.  Even if the heavy elements
are removed from early galaxies by strong winds, there are constraints
on the distance that these winds will cover based simply on travel
time (e.g. $D\sim 10$ kpc per 100Myr for a 100 \kms wind).  It is also
likely that early galactic winds would stall at $R\lesssim 100$ kpc
because of mass loading from both the ISM and the IGM, which is
$300-1000$ times more dense at these redshifts than in the present
day.

To date, no abundance decline has been observed in the IGM towards
high redshifts.  Over the $z\sim 2-4$ range where one can estimate
ionization-corrected carbon abundances from \civ,
\citet{schaye_civ_pixels} found no significant redshift evolution in [C/H]
across a sample of several sightlines.  Likewise,
\citet{songaila_omegaz, songaila_new_civ} and \citet{pettini_z5_civ}
found no evolution in $\Omega_{\mciv}$, the integrated contribution of
\civ gas to closure density, between $z=1.8$ and $z=5$.  The latter
result is particularly surprising since $\ociv$ depends on both the
overall carbon abundance and the \civ ionization fraction
$n_\mciv/n_{\rm C}$.  Naively one would expect both of these
quantities to vary over such a broad redshift range, because of
ongoing chemical enrichment and the evolution of the ionizing
background radiation spectrum.

Despite the discovery of several $z\gtrsim 6$ QSOs in the last few
years \citep{fan_z6qsos_paper1, fan_z6qsos_paper2,
fan_z6qsos_paper3,fan_z6qsos_paper4}, intergalactic absorption-line
studies have been limited to $z\lesssim 5$ because at higher redshift
the \civ doublet moves into the near-infrared.  Absorption line
spectroscopy is much more challenging in this regime because of
increased detector noise, OH emission from the sky, and more severe
telluric absorption.  These foreground problems can be mitigated by
employing higher spectral resolution, but the associated penalties in
sensitivity require long integrations to overcome.

In this paper I present $J$ band GNIRS spectra of two $z\sim 6$ QSOs
at moderately high spectral resolution ($\Delta v=60$ km/s).  These
spectra are searched for \civ absorption systems at $5.4\lesssim z
\lesssim 6.2$, and the results used to extend existing measurements
of $\ociv$ to $z\sim 6$.  In section \ref{sec:observations}, I describe
the observations and reduction methods.  Section \ref{sec:analysis}
details the \civ search procedure and simulations to quantify the
effects of incompleteness and/or spurious \civ detections in the data.
Sections \ref{sec:results} and \ref{sec:discussion} present the
results for $\ociv$ and a discussion of their implications and
limitations.

\section{Observations}\label{sec:observations}

\begin{figure}
\epsscale{1.25}
\plotone{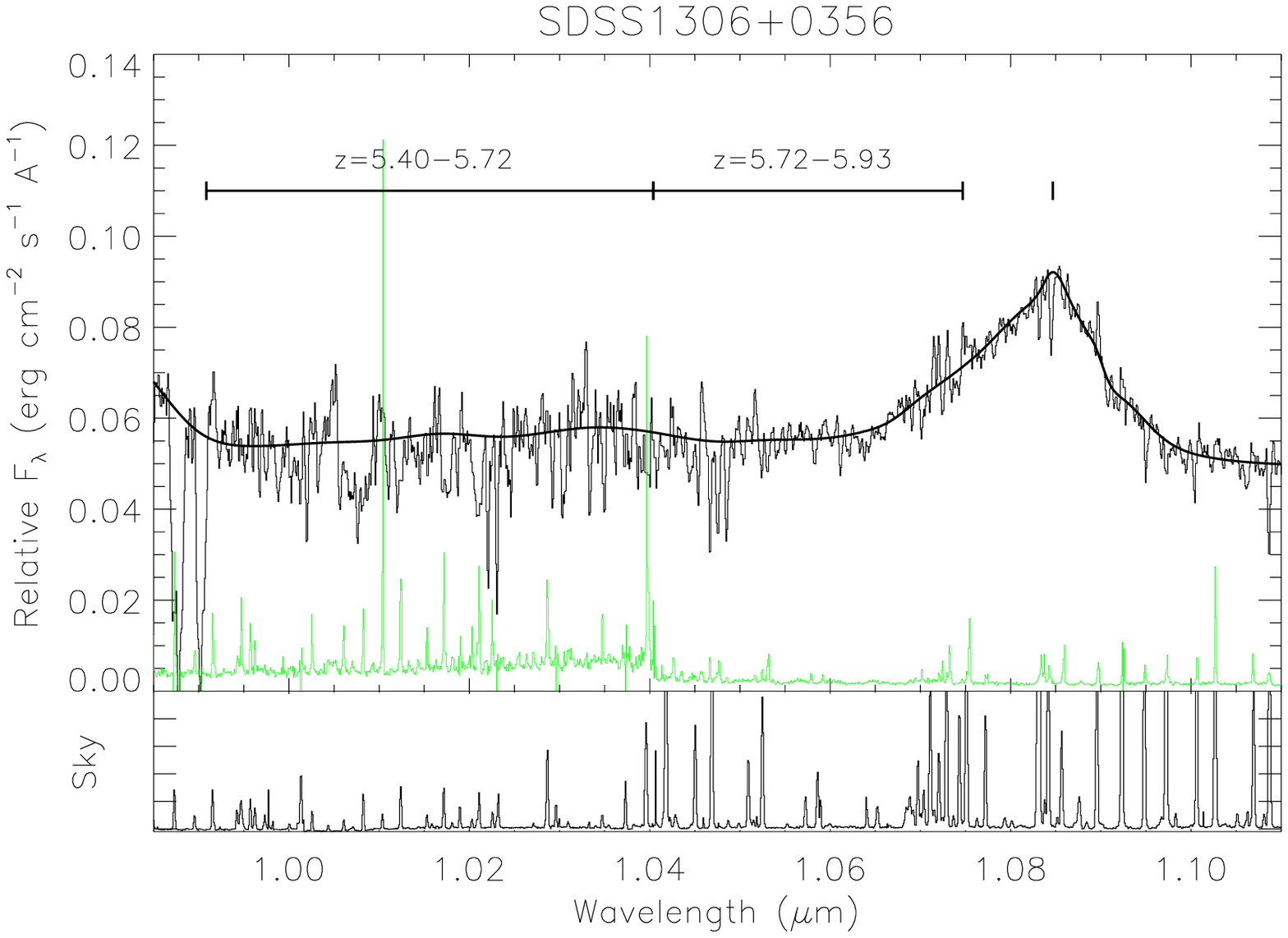}
\vskip 0.1in
\plotone{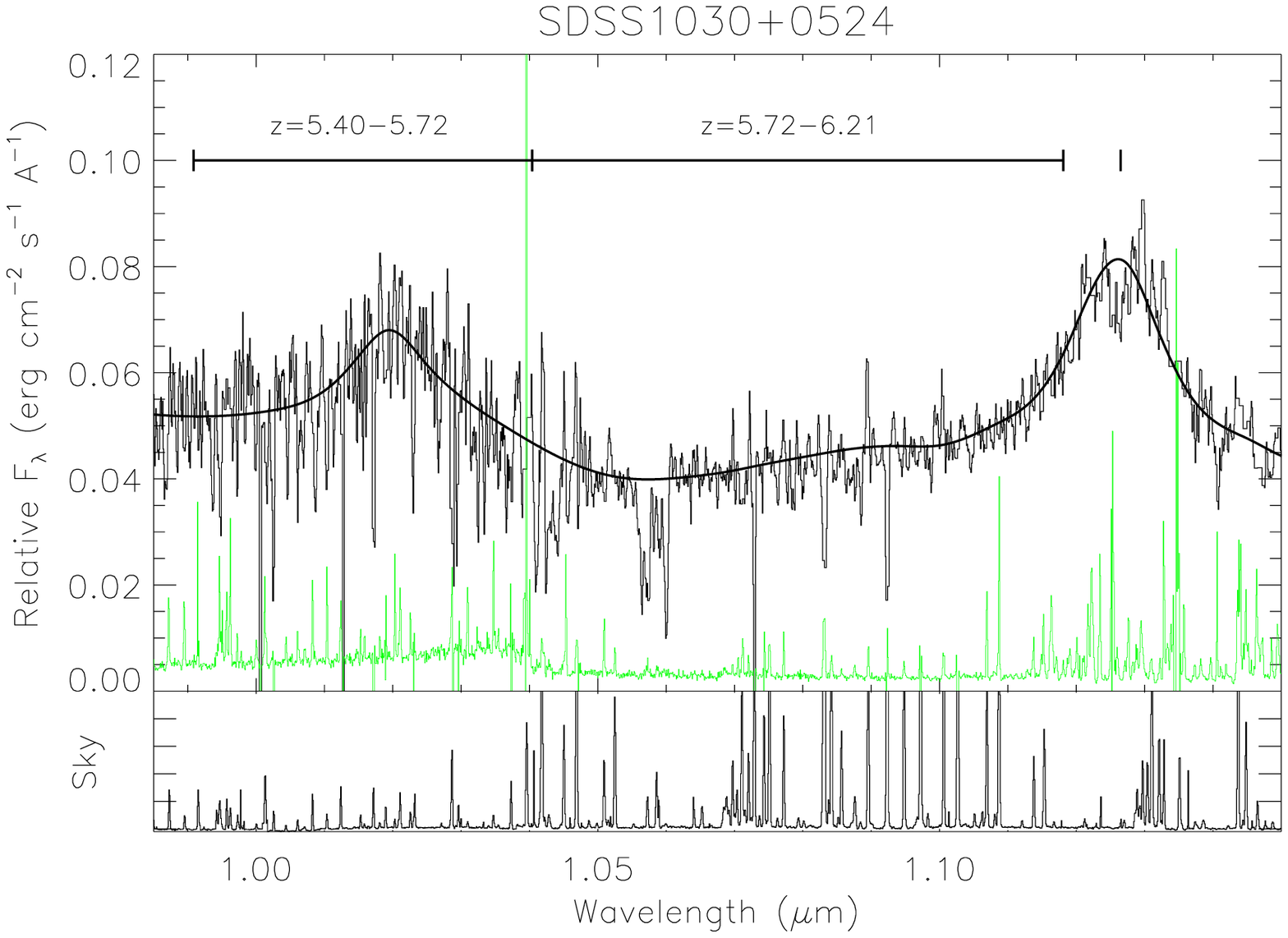}
\caption{\civ spectra of $z\sim 6$ QSOs SDSS1306+0356 ($z_{em}=6.002$)
and SDSS1030+0524 ($z_{em}=6.272$).  Solid line shows continuum
estimate, and light green line shows $1\sigma$ error estimates,
including the effects of Poisson noise from sky emission and telluric
absorption.  Night Sky emission is shown in bottom panels, and the
\civ search ranges are indicated with labeled horizontal bars.  The
redshift of the \civ emission peak is shown with a vertical dash.}
\label{fig:spectra}
\end{figure}

Figure \ref{fig:spectra} displays the \civ spectral regions for the
high redshift QSOs SDSS1306+0356 ($z_{\rm em}=6.002$ as measured from
the \civ emission peak) and SDSS1030+0524 ($z_{\rm em}=6.272$).  The
data were obtained with the GNIRS instrument \citep{elias_gnirs},
operated in queue mode on Gemini-South during semesters 2005A and
2006A.  The instrument was configured with the cross-dispersion prism,
the 110 line/mm grating, and a 0.45 arcsecond slit.  Two grating
positions were observed for each QSO to provide complete spectral
coverage up to the emission wavelength of \civ, although the
signal-to-noise ratio (SNR) is not completely uniform between setups
due to varying conditions on different nights within the queue.  Each
QSO was observed for approximately 8 hours per setup.

The observations were split into sequences of $4\times10$ minute
integrations, nodded four arcseconds along the slit in an ABBA
configuration.  The data were reduced using customized IDL software
developed by the author.  Individual frames were filtered for detector
pattern noise and flatfielded, and the ABBA frames were combined to
remove signatures of the sky, dark current, and charge persistence.
Each 2D pixel was assigned a wavelength by fitting a second order
polynomial to the centroids of OH sky lines, which were compared with
the line lists compiled by \citet{rousselot_oh}.  Residual errors in
the sky subtraction were corrected by a second-pass fit of a b-spline
function to the subsampled grid of wavelength versus pixel intensity.
Objects were identified and extracted using an optimal Gaussian
profile, with errors determined by direct calculation of the sky RMS
outside of the object profile (the errors were not dominated by shot
noise from the object).  The data were flux calibrated and corrected
for telluric absorption using observations of standard stars obtained
throughout the night at similar airmass as the QSOs.  Finally, the
extracted spectra were scaled to match in flux, and added together on
a 1D grid of pixels with constant velocity spacing of $25.3$ km/s.
Exposures from different nights were weighted according to their
$({\rm SNR})^2$, and the wavelengths converted to vacuum heliocentric
coordinates.  The final resolution of the combined spectra was
measured as $R=5000$ or $60$ km/s FWHM (2.35 pixels) using Gaussian
fits of sky lines in both the reduced 1D and the raw 2D frames.

\section{Analysis}\label{sec:analysis}

The data in Figure \ref{fig:spectra} are among the first $z>5$ IR QSO
spectra with resolution suitable for absorption line measurements;
however, they are still a factor of $\sim 10$ coarser than high
resolution optical spectra at $z\sim 3$ and have lower SNR by a factor
of several.  At $\Delta v=60$ km/s per resolution element, they are
sufficient to resolve the \civ doublet (with characteristic $\Delta
v\sim 500$ km/s).  However, individual \civ components at lower
redshift have typical intrinsic velocity dispersions of $b\sim 10$
km/s \citep{rauch_civ_kinematics}, and hence will generally not be
resolved in these spectra.  Since the line broadening is dominated by
instrumental effects, I have characterized the absorption systems
through a curve-of-growth analysis with attention to possible
saturation effects, rather than by direct Voigt profile fitting.
However, all of the results were compared against Voigt profiles
generated using independent software as a final consistency check.

The data were divided into two separate redshift intervals for the
analysis, with the interval boundaries determined by the SNR of the
data.  The redshift ranges are indicated in Figure \ref{fig:spectra}
by horizontal bars.  In both objects an increase in the pixel noise
level is seen at the transition between two spectral orders at
$\lambda\approx1.04~\mu$m.  This established the lower boundary of the
high-redshift bin at $z=5.72$.  The upper boundary of the
high-redshift bin was set at $z=6.21$, or $2500$ \kms ~below the
emission redshift of SDSS1030+0524.  It is common practice in
absorption line surveys to omit this last portion of spectral coverage
because any absorbers there would be affected by local radiation from
the QSO.  For this same reason the sarch range of SDSS1306+0356 was
limited to $z=5.72-5.93$ for the high redshift bin.  Neither
SDSS1030+0524 or SDSS1306+0356 contains any \civ lines within $2500$
\kms ~of its emission redshift, so the results are not sensitive to
this choice.  Inclusion of the extra spectrum would amount to a $\sim
15\%$ ($0.06$ dex) increase in pathlength.

Because of its lower emission redshift, SDSS1306 does not cover the
entire range of the high redshift bin.  Hence the effective
(pathlength weighted) central redshift of the bin falls at $z=5.92$,
slightly below the nominal midpoint of $z=5.97$.  The low redshift bin
ranges from $z=5.40-5.72$, with its lower bound chosen to avoid
contamination from interloping \mgii at $\lambda\approx 9900\AA$ in
SDSS1306+0356.  Both objects cover its full width.

Because of the IR data's lower spectral resolution, lower SNR, and
shorter pathlength relative to $z\sim 3$ optical surveys, this initial
abundance measurement at $z\sim 6$ will not be sensitive to weak \civ
systems in the very tenuous IGM.  However, the data quality is high
enough to detect the stronger \civ systems commonly seen in lower
redshift data with high confidence.  Because the slope of the \civ
column density distribution is flatter than $N_\mciv^{-2}$
\citep{songaila_omegaz}, it is these systems which provide the
dominant contribution to $\ociv$ \citep{pettini_z5_civ,
simcoe_feedback}, and this is the main measurement presented in
Section \ref{sec:omega_civ}.

\subsection{Identification and Measurement of {\rm \civ}Lines}

Each of the QSO spectra was searched systematically for \civ doublets
using an automated software routine.  First, an estimate of the
continuum was constructed by interpolating an interactively fit cubic
spline.  The minimum spacing of the spline knots was set at $500$ \kms
~to avoid overfitting the continuum on smaller scales.

\begin{figure}
\epsscale{1.20}
\plotone{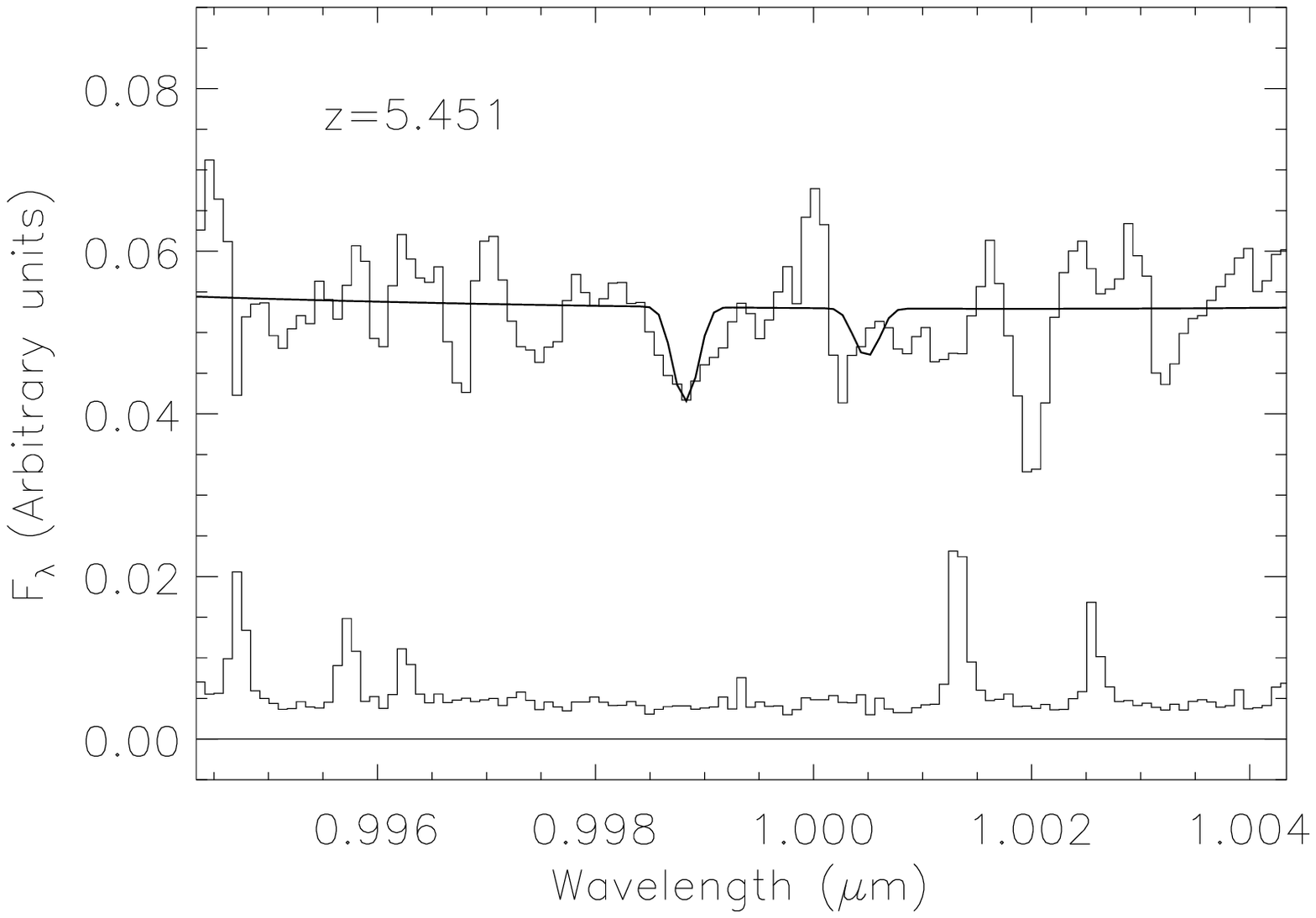}
\plotone{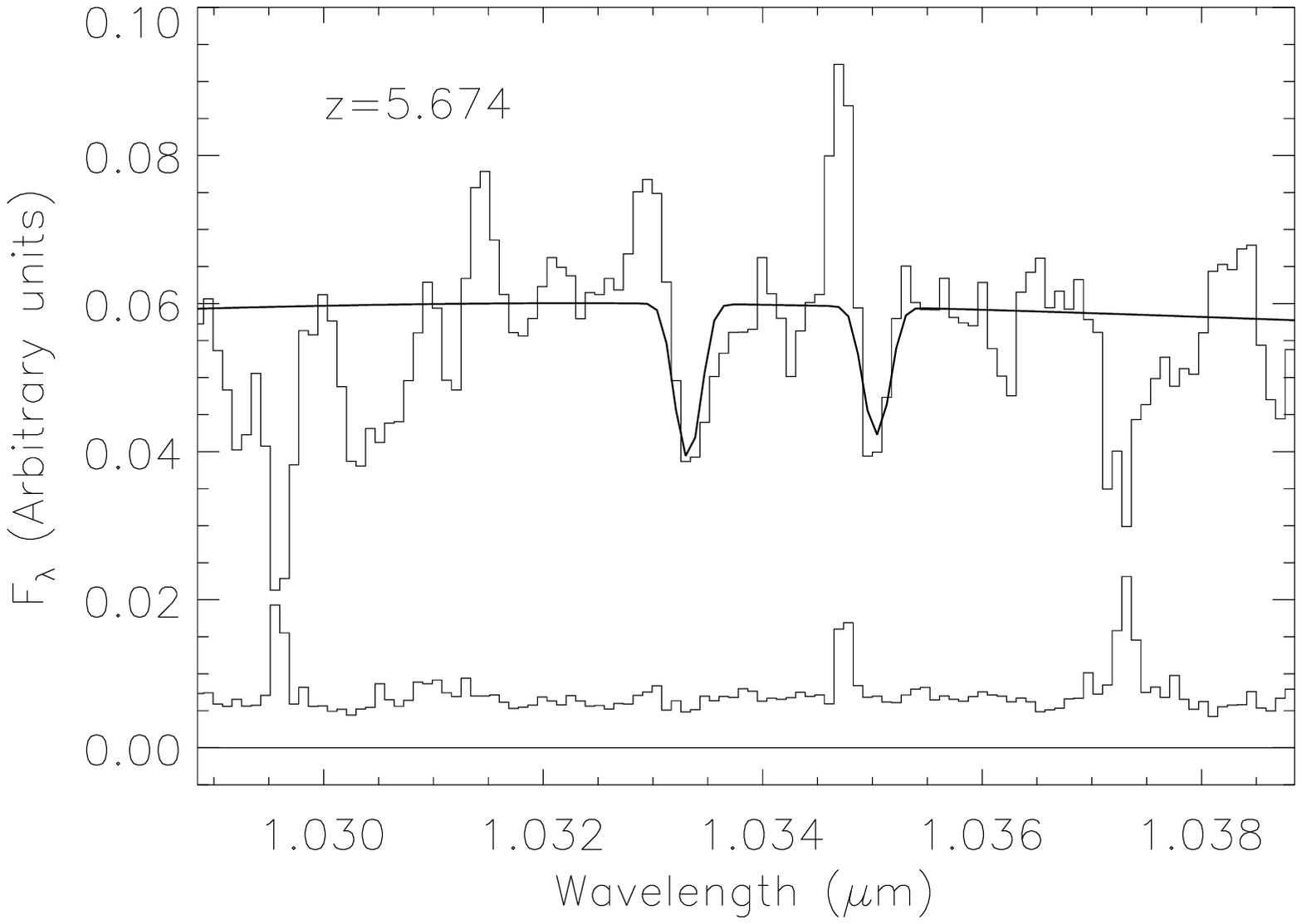}
\plotone{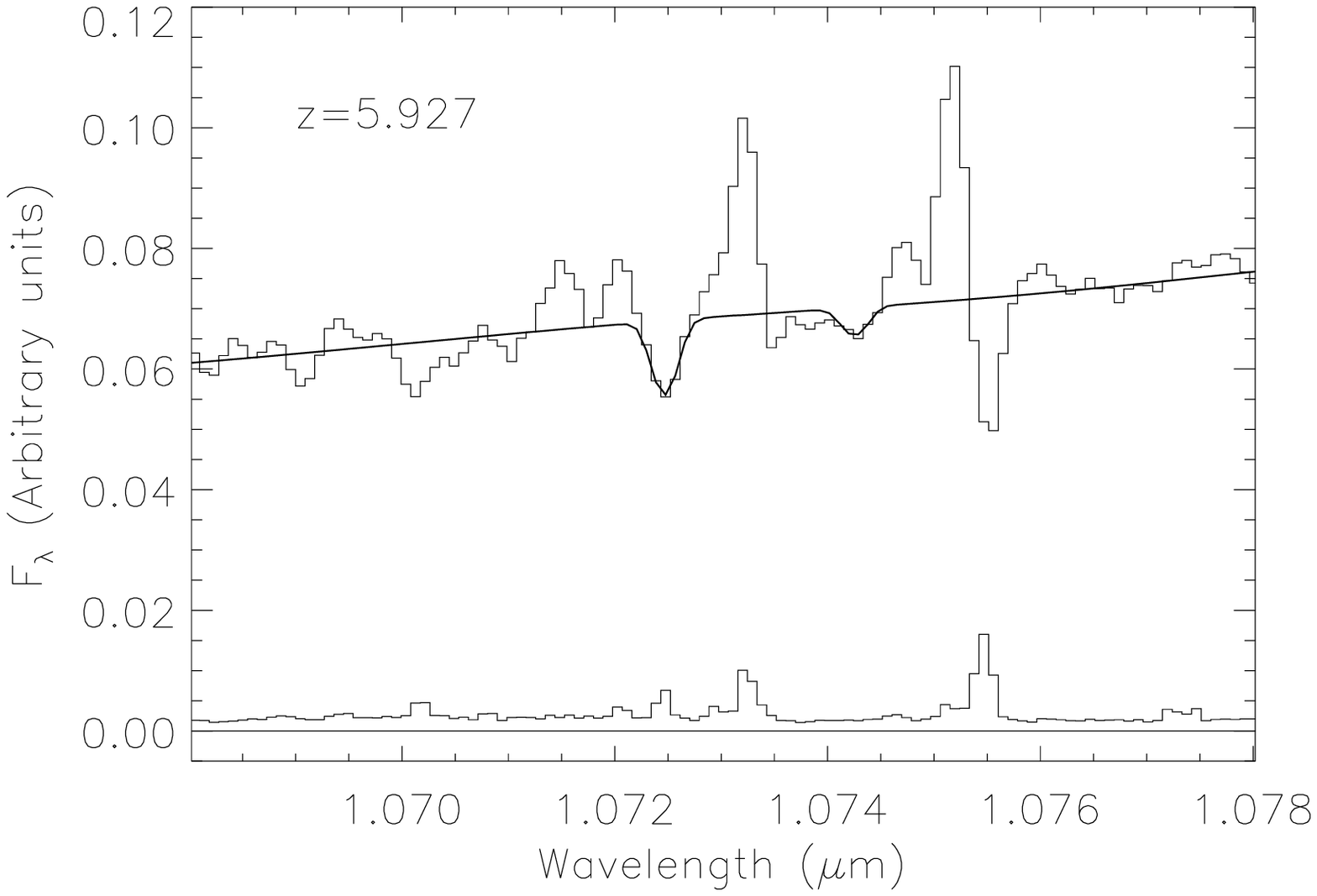}
\caption{\civ detections along the line of sight towards
SDSS1306+0356.  Best-fit absorption models are shown with solid lines.
The upper two systems are contained in the low-redshift sample; the
bottom system is a marginal detection, and it is the only \civ line
identified at high redshift toward this QSO.}
\label{fig:sdss1306_sys}
\end{figure}

The entire redshift range was then scanned for absorption feaures with
$\ge 3\sigma$ significance in equivalent width (EW) over a resolution
element.  Each flagged feature was considered a possible $1548.2$\AA
~component of the \civ doublet, and the EW was then calculated for the
corresponding $1550.8$\AA ~region.  If the 1550.8\AA ~EW was also $\ge
2\sigma$, the system was flagged as a candidate \civ doublet.  The
list of candidate doublets was then examined by hand to discard
systems that were obviously contaminated by sky subtraction residuals
or noise, or which had doublet ratios inconsistent with \civ
(i.e. $W_{1551} > W_{1548}$).

\begin{figure}
\epsscale{1.1}
\plotone{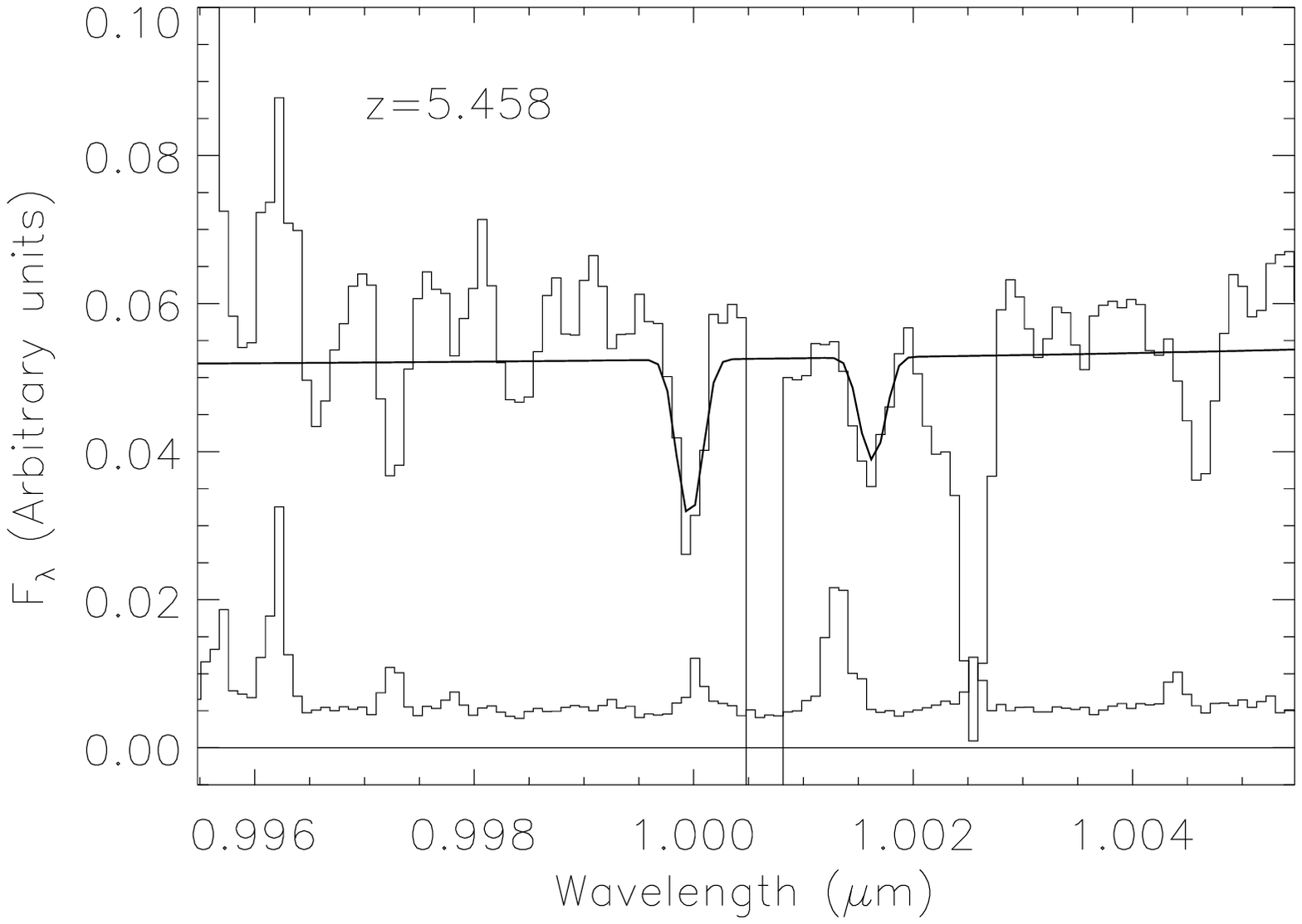}
\plotone{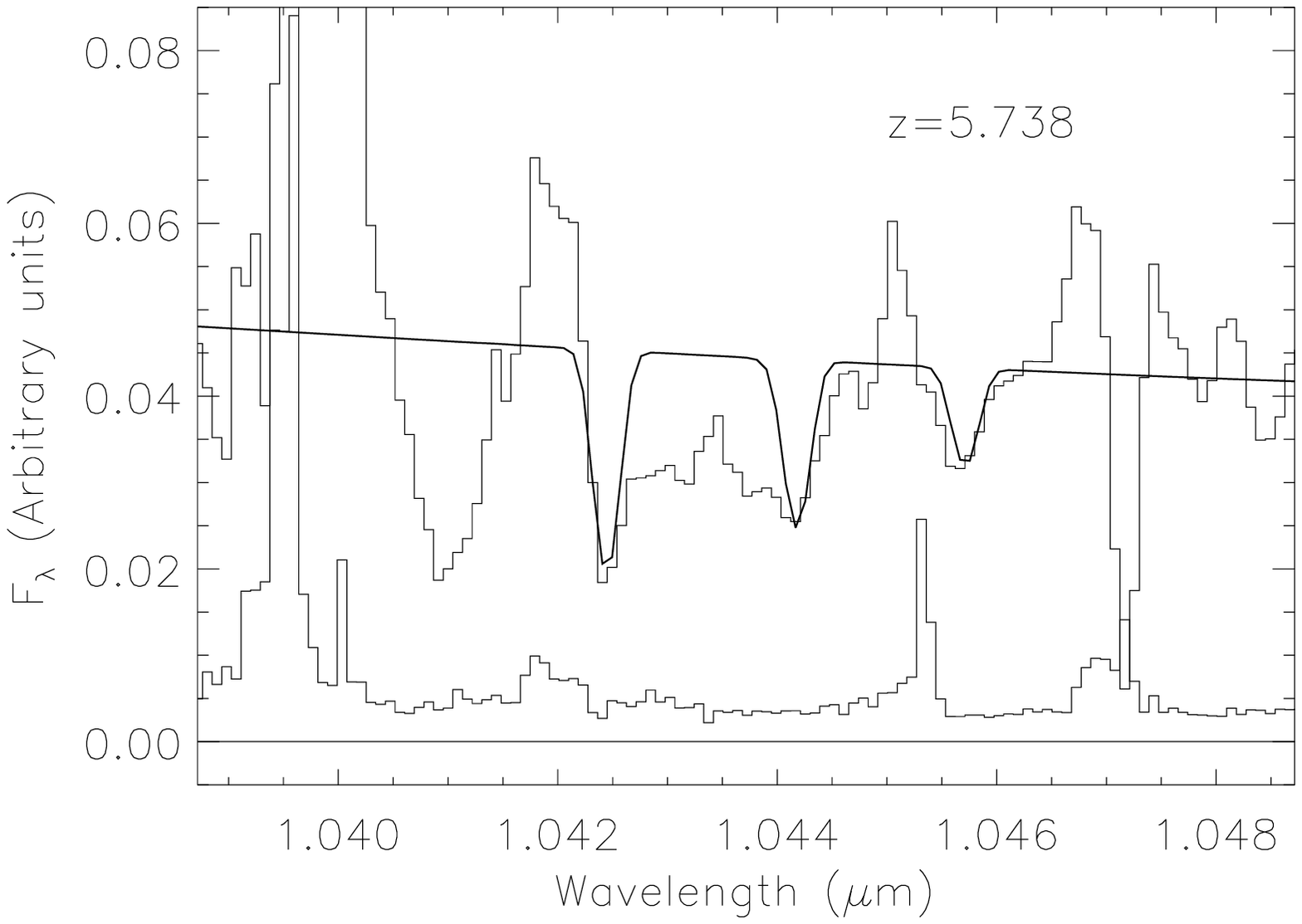}
\plotone{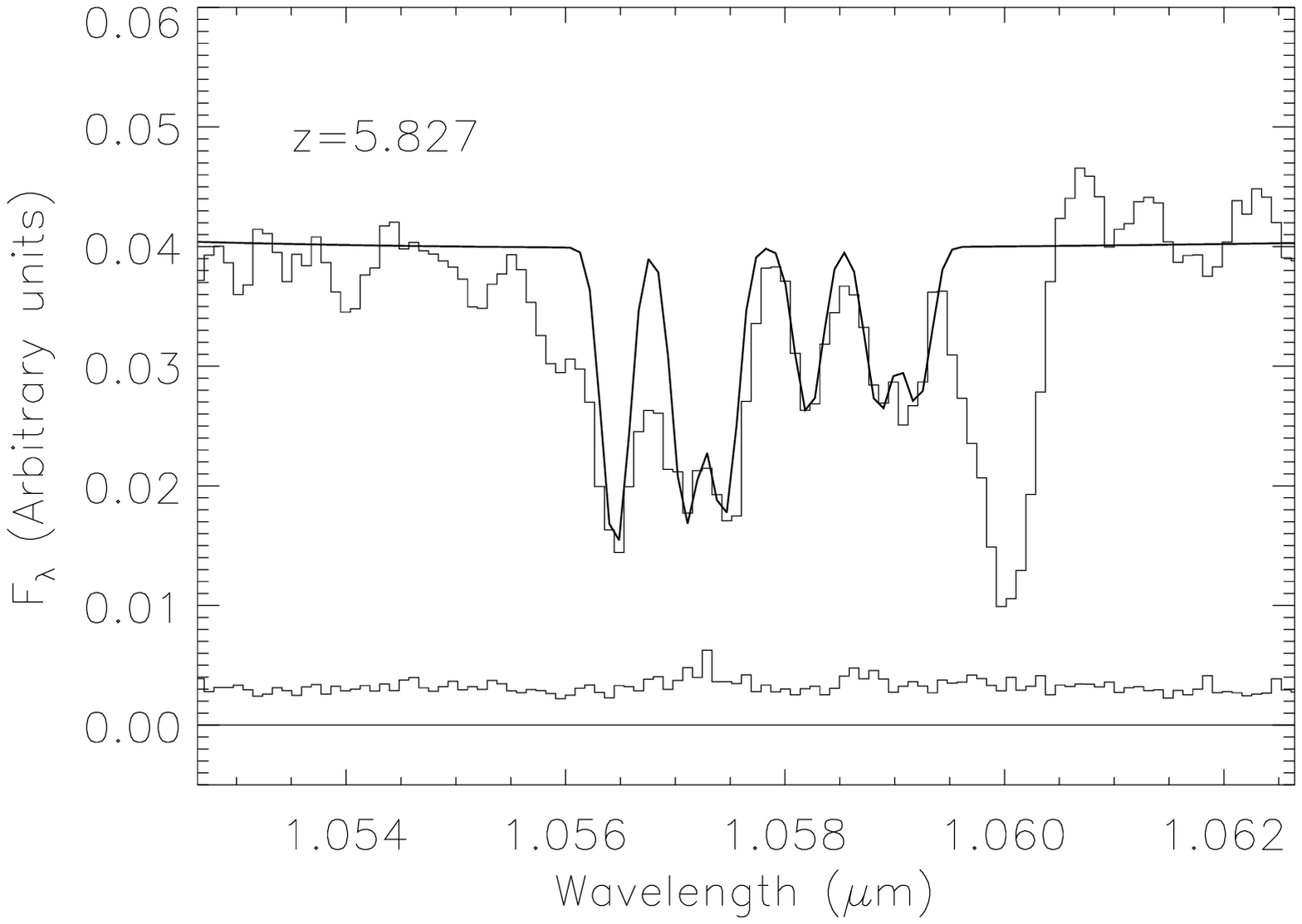}
\plotone{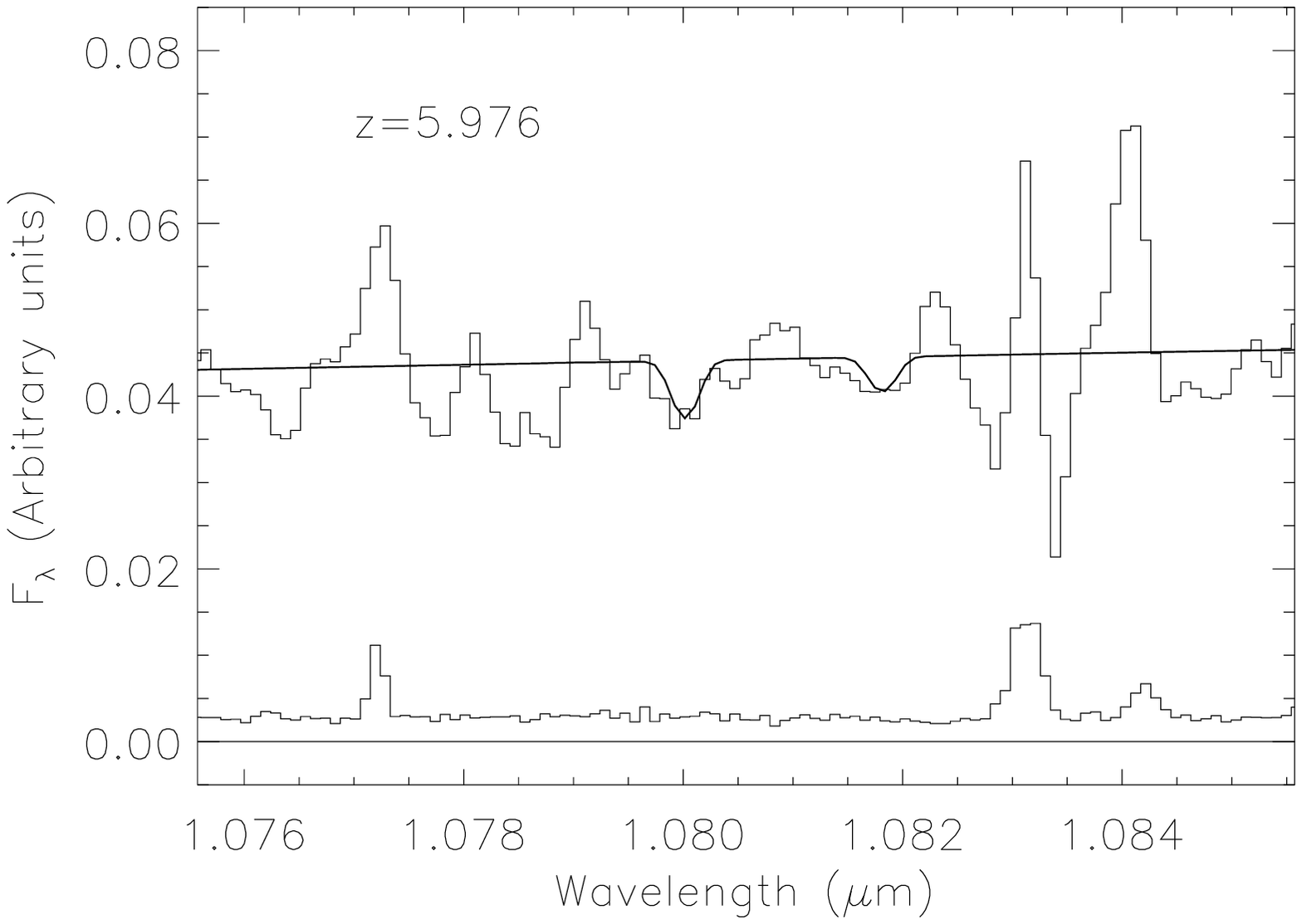}
\caption{\civ detections along the line of sight towards
SDSS1030+0526.  Top system at $z=5.458$ is in the low redshift sample,
all others are in the high redshift sample.  Best-fit absorption
models are shown with solid lines.}
\label{fig:sdss1030_sys}
\end{figure}

\begin{figure}
\epsscale{1.1}
\plotone{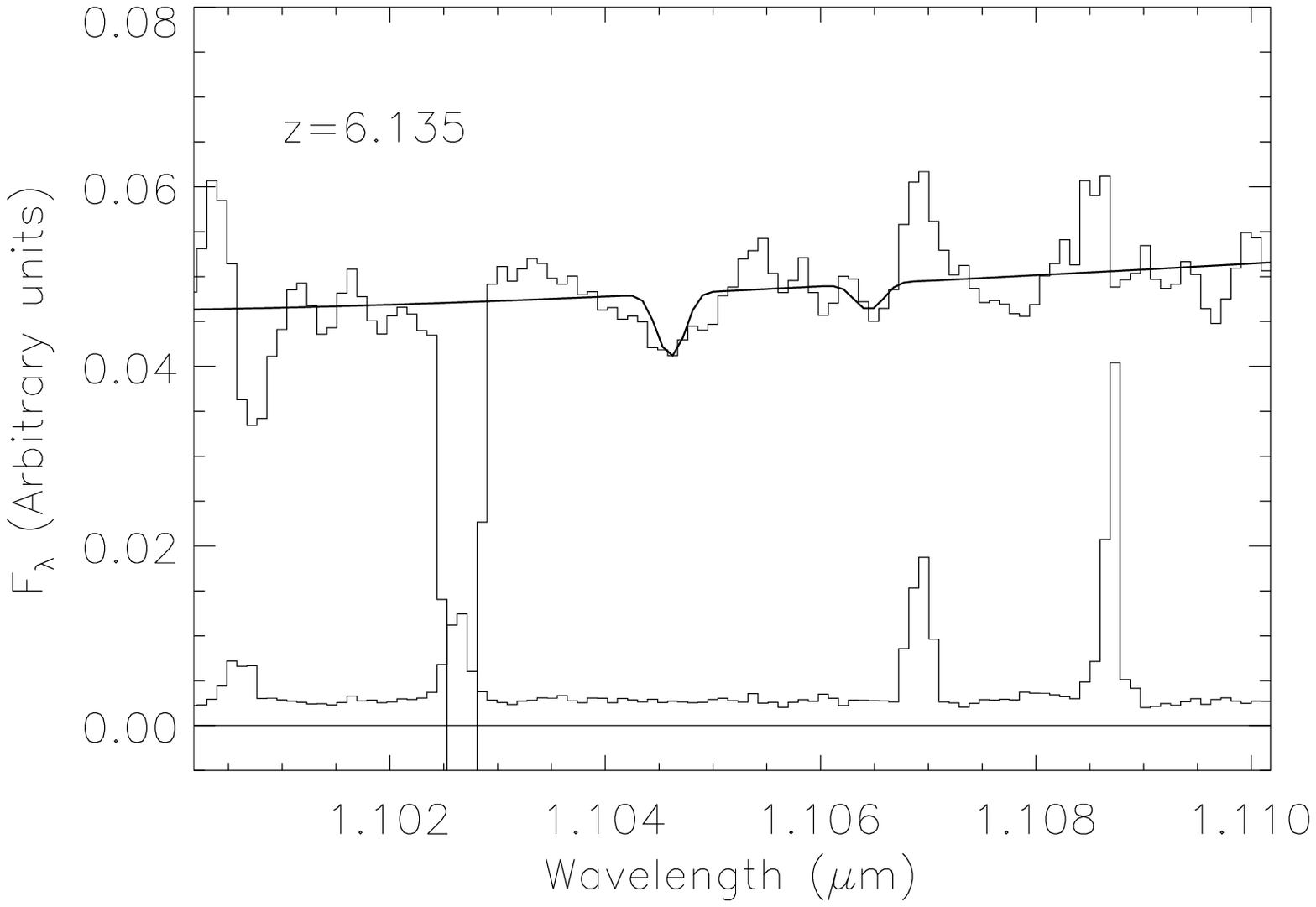}
\plotone{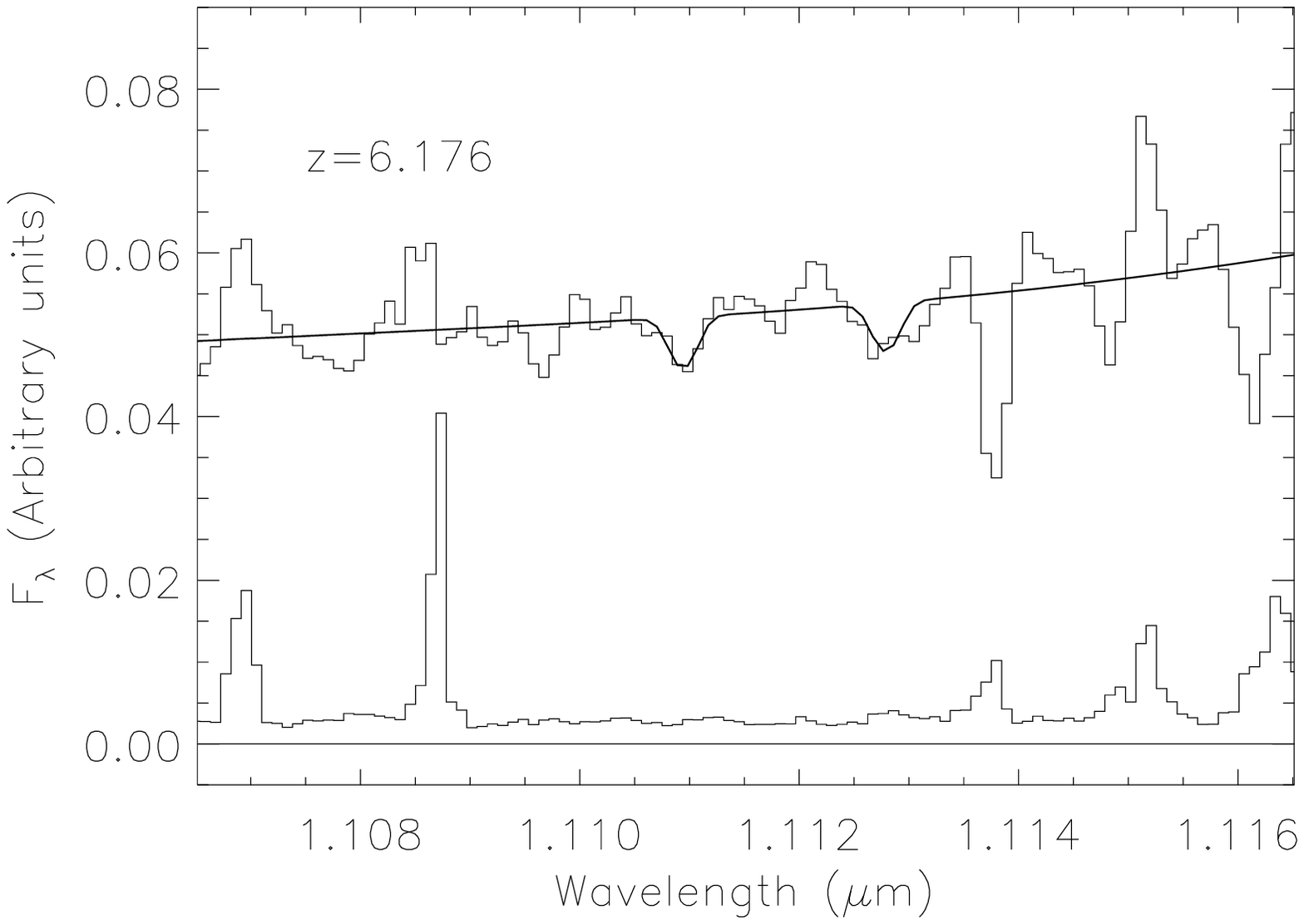}
\caption{\civ detections along the line of sight towards SDSS1030+0526
(Continued from previous figure).}
\label{fig:sdss1030_sys2}
\end{figure}

Application of this procedure resulted in nine \civ detections between
the two sightlines, ranging from $z=5.451$ to $z=6.175$.  For each of
these systems, I estimated a more accurate rest EW of \civ ($W_{\rm r,
\mciv}$) by fitting a Gaussian absorption profile to the data, with
its width fixed at one resolution element.  Plots of each system and
its best-fit absorption profile are shown in Figures
\ref{fig:sdss1306_sys} through \ref{fig:sdss1030_sys2}, with absorption
line redshifts and EWs listed in Tables 1 and 2.

Some of the high redshift \civ lines are highly significant
detections, while some are more marginal.  Along the SDSS1306
sightline (which has a higher SNR) only three systems are present.
The system at $z=5.674$ is a strong detection, although it is blended
with noise features in the data.  The systems at $z=5.451$ and
$z=5.927$ are much weaker and even with a high SNR over these portions
of the spectrum, the \civ $1550.8$\AA ~detection significance is
low.

Toward SDSS1030+0524 the search algorithm identified six \civ systems.
Two strong detections are found at $z=5.738\pm0.005$ and
$z=5.827\pm0.005$.  The system at $z=5.738$ contains two significant
components, with the $1548.2$\AA ~component of one line blended with
the $1550.8$\AA ~component of the other.  It is found in a noisy
region of the spectrum, where systematic errors in the sky subtraction
are non-negligible.  The $z=5.827$ system contains three highly
significant doublets.  This complex also contains another strong line
to the red which could not be identified with any ion commonly seen in
QSO spectra at any redshift.  All well-known QSO lines with rest
wavelengths between 1450\AA ~(corresponding to the QSO's rest frame)
and 6000\AA ~(i.e. past the NaD doublet) were considered without
finding a match.  The presence of this absorption was verified in the
2D trace over multiple nights, and checked to rule out artifacts from
telluric absorption or features in the flat field.  It was also
checked against a high-resolution optical spectrum of SDSS1030 (kindly
provided by G. Becker) without resulting in any cross-identification.
The most likely interpretation appears to be a mixture of three \civ
components with one additional unidentified line.

Only two weak \civ systems are identified at $z>6$.  They are each
near the detection limit of the data, and both are found in relatively
clean and sky-free portions of the spectra.  It is intriguing that the
strong \civ systems generally seem to be found at $z\lesssim 5.9$,
while at $z\gtrsim 5.9$ (where the SNR of our data is highest) the
\civ systems are weaker.  While this could be consistent with some
evolution of the population, this effect could also be caused by shot
noise along a short redshift path, as described in more detail in
Section \ref{sec:cddf}.

\begin{figure}
\epsscale{1.25}
\plotone{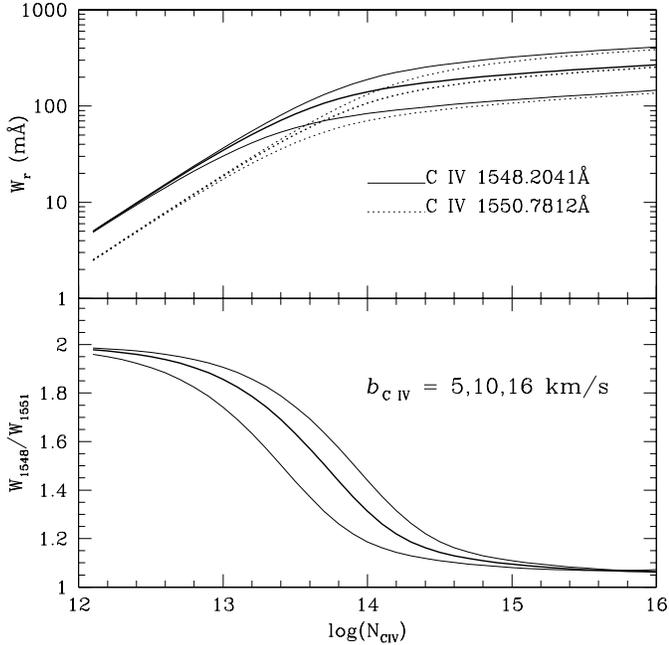}
\caption{Curve-of-growth calculations for \civ.  Top panel shows the
relation of $W_r$ to $\nciv$ for both components of the doublet.
Bottom panels shows the equivalent width ratio of the two doublet
components as a function of $\nciv$, for three typcal choices of $b$
parameter.}  
\label{fig:cog}
\end{figure}

Some, but not all, of the \civ doublets show evidence of
instrumentally unresolved saturation in the physical absorption
profile.  Figure \ref{fig:cog} shows the \civ curve-of-growth (COG) of
rest EW ($W_r$) versus $\nciv$; the bottom panel displays the EW ratio
between the two components as a function of $\nciv$ for various values
of the \civ $b$ parameter.  For reasonable values of $b_{\mciv}$, the
1548.2\AA ~ component begins to saturate near $\nciv\sim 10^{13}$
\pcmsq, and the doublet ratio approaches unity as the 1550.8\AA
~component also begins to saturate at $\nciv\sim 10^{14}$ \pcmsq.

Column 7 in Tables \ref{tab:sdss1306} and \ref{tab:sdss1030} shows the
measured doublet ratio for each \civ detection.  Because of the data's
sensitivity limitations, most of the detected systems lie in the
non-linear regime between $10^{13}\lesssim\nciv\lesssim10^{14}$
\pcmsq, where the $\nciv-W_r$ mapping is degenerate with the \civ $b$
parameter.  However, even for some of the stronger systems in the data
with $W_r\sim 200-300$ m\AA ~(e.g. the complex at $z=5.823-5.830$),
the doublet ratios provide useful upper bounds on $\nciv$.  Often the
ratio is larger than one would guess from measuring $W_r(1548)$ alone,
and in these cases it is likely that the observed \civ line is
actually an unresolved blend of several narrow components with more
complex velocity structure.

At $z\sim 3$, the typical \civ line has a velocity dispersion of
$b\approx 10$ \kms, and over $80\%$ of \civ components have $b_\mciv
\le 16$ \kms \citep{rauch_civ_kinematics}.  Assuming $b_\mciv=10$ \kms
~for our systems, the doublet ratio can be used to estimate $\nciv$.
Then, the number of unresolved components can be inferred by comparing
the measured value of $W_r$ with the single-component value from the
COG.  The rightmost columns in Tables \ref{tab:sdss1306} and
\ref{tab:sdss1030} show the $\nciv$ value deduced from the EW ratios,
along with the corresponding number of components required to match
the total EW for each system.

The $\nciv$ values for individual components range between
$10^{13.15}$ and $10^{14.3}$.  Different choices of $b_\mciv$ can
change specific values of $\nciv$ by up to $\sim 0.2$ dex; random
measurement errors in the EW could add another 0.2 dex of uncertainty,
though these errors would be uncorrelated.  However, the {\em total}
column density of each \civ system is fairly insensitive to the
specific choice of $b$.  A larger $b$ parameter leads to fewer
unresolved components with higher $\nciv$, while a small $b$ leads to
more components with smaller column densities, but the sum of the
columns matches to within $<0.1$ dex for any choice of $b$ between 5
and 20 \kms.

\subsection{Estimates of Sample Completeness}\label{sec:completeness}

IR absorption line searches require particular attention to
characterize sample incompleteness and false-positive identifications.
In addition to normal SNR variations in the data, one encounters many
spectral regions where Poisson noise from OH sky emission or residuals
from imperfect telluric absorption correction render pixels unusable
for cosmological absorption measurements.  At $R=5000$, the GNIRS
spectra resolve sky features well enough to work in many of the dark
regions between fetures from the night sky.

To quantify the completeness of the $z\sim 6$ \civ sample as a
function of $\nciv$, we have performed a series of Monte Carlo
simulations.  For each redshift bin in each sightline, a sample of
artifical \civ doublets was created with $\nciv$ ranging from
$10^{12}-10^{16}$ \pcmsq ~and redshifts selected at random.  Each
artifical line was added individually into the real spectrum, and the
data were searched using the same EW criteria to see if the artificial
line was recovered.  One thousand trials were performed for each
column density.

\begin{figure}
\plotone{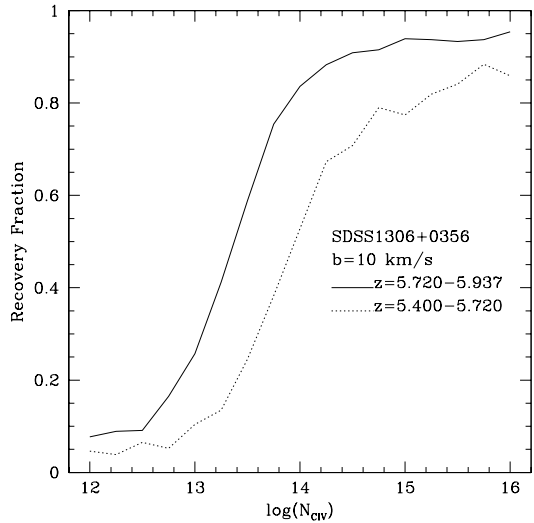}
\plotone{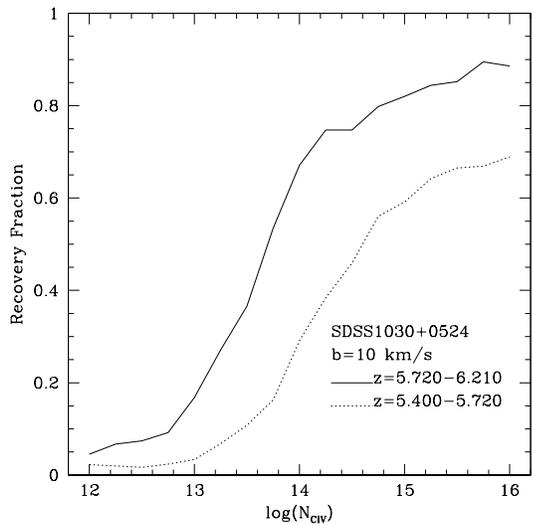}
\caption{Estimates of sample completeness as a function of $N_\mciv$,
determined from Monte Carlo simulations performed on the data.  Top
panel shows completeness for SDSS1306 over both sample redshift bins;
bottom panel shows the same for SDSS1030.  \civ lines are recovered
more completely in the higher redshift bins because of improved SNR in
these spectral regions.}
\label{fig:completeness}
\end{figure}

Figure \ref{fig:completeness} displays $f_{\rm comp}$, the fraction of
lines recovered as a function of $\nciv$ for each of the sightlines.
For most of the redshift path, the data are largely complete ($\gtrsim
70\%$) at $\nciv \gtrsim 10^{14}$ \pcmsq.  The completeness drops off
rapidly towards $\nciv \le 10^{13}$ \pcmsq, at which point only
$5-10\%$ of systems are detected.  In the IR, unlike the optical, it
is nearly impossible to achieve 100\% completeness even at thie
highest column densities because of chance coincidences with OH sky
lines or strong telluric features.  However, for the \civ lines at
$\nciv\gtrsim 10^{14}$\pcmsq, it appears that at $R\sim 5000$ one
resolves night sky foregrounds out of $\sim 80\%$ of the total
available pathlength.

Sample incompleteness effectively reduces the absorption pathlength
over which lines of a given column density can be detected in the
survey.  In the discussion below, we explicity state when completeness
corrections are applied to the data.  These are implemented by scaling
the relevant pathlengths $\Delta X$ by the factors shown in Figure
\ref{fig:completeness}.

\subsection{Incidence of False Positive Identifications}\label{sec:falsepos}

Several of the \civ detections in Tables \ref{tab:sdss1306} and
\ref{tab:sdss1030} approach the sensitivity limit of the data, and at
these column densities one must also consider the possibility of false
positive identifications from chance noise fluctuations.  This process
is much more difficult to simulate since in true spectra the noise
properties are often not perfectly approximated by a Poisson
distribution, especially in the very instances where correlated pixels
lead to false detections.

One method for quantifying the false positive rate is to scan the
spectra for absorption ``doublets'' identical to \civ in every sense,
except that the oscillator strengths of the two transitions are
reversed.  Systems identified this way will be subject to the same
noise properties and line spacing as true C~{\sc iv}, but can only be
caused by correlated noise fluctuations in the data since a doublet
with $W_r(1551)>W_r(1548)$ is unphysical.

I searched for this type of system in the two QSO spectra using the
same EW criteria as for the true \civ search.  For SDSS1306, the high
redshift sample at $z=5.72-5.93$ yielded one weak ``false'' system.
For SDSS1030, there was one (rather strong) flase system at $z=5.722$,
and two other weak systems at $z=6.005$ and $6.155$.  Over the full
$z=5.72-6.21$ redshift path, the total column density of all ``true''
\civ systems is $10^{14.8}$ \pcmsq, whereas the the total for the
``false'' systems was $10^{14.3}$ \pcmsq.  Contamination from false
positives may therefore comprise up to $\lesssim 30\%$ of the reported
\civ in this bin, but spurious detections do not appear to dominate
over the true \civ signal.  In the $z=5.40-5.72$ bin the false
positive signal is likewise estimated at $\lesssim 25\%$.
%

If one restricts the search to the $z\gtrsim 6$ region, the
contribution from contaminants is no longer subdominant.  The
integrated \civ column for the three ``true'' \civ systems is
$\sum\nciv = 10^{13.6}$ \pcmsq, whereas the two ``false'' systems have
$\sum\nciv = 10^{13.7}$.  This result is consistent with the
interpretation that {\em all} of our $z>6$ \civ detections are
spurious, or alternatively that there is no statistically significant
signal from \civ over $z=6.0-6.21$ in SDSS1030.  This is an intriguing
possibility, but the absorption pathlength is fairly short for this
one sightline, so it could just be the result of statistical
fluctuations in a small sample.  I return to this point in the
discussion; for the purpose of calculations below I continue to treat
the $z>6$ \civ detections as real, keeping in mind that because of
contamination they will lead to upper limits on the values of any
derived quantities.

\section{Results}\label{sec:results}

\subsection{$z\sim 6$ \civ Column Density Distribution}\label{sec:cddf}

\begin{figure}
\epsscale{1.2}
\plotone{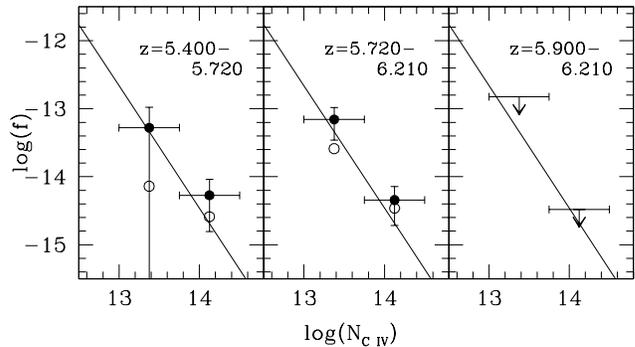}
\caption{The \civ column density distribution function, defined as
$f(\nciv)=d\mathcal{N}/d\nciv dX$, over three different redshift
intervals.  Points are shown with incompleteness corrections; the
uncorrected values are shown with open circles.  Left and center
panels show the measured CDDF for the nominally defined redshift bins.
The right panel illustrates limits for a possible high redshift bin,
where more absorption pathlength would be required to obtain an
accurate measurement.}
\label{fig:cddf}
\end{figure} 

A key result from lower redshift \civ surveys
\citep[e.g.][]{songaila_omegaz} is that the column density
distribution function (CDDF) of \civ---defined as the number of
absorbers per unit $\nciv$ per unit absorption pathlength---is nearly
invariant between $z\sim 2$ and $z\sim 5$.  Even with the small sample
available at $z\sim 6$, one can construct a coarse CDDF for comparison
with this lower redshift data.

%
For calculating the absorption pathlength, I use the convention that
$\Delta X = X(z_2)-X(z_1)$, where $X(z)$ is defined as
\begin{equation}
X(z) = {{2}\over{3\Omega_M}}\sqrt{\Omega_M(1+z)^3+\Omega_\Lambda}
\label{eqn:pathlength}
\end{equation}
with $\Omega_M=0.3$ and $\Omega_\Lambda=0.7$.  Results from the
literature are often quoted using the $\Delta X$ for an
Einstein-deSitter universe; where appropriate I have rescaled the
literature values by $\sqrt{\Omega_M}$ to conform with more recent
concordance models.  

Given the small size of the $z\sim 6$ line sample, I only estimate the
CDDF in two $\nciv$ bins---one ranging from $\log(\nciv)=13.00-13.75$
and one from $\log(\nciv)=13.75-14.5$.  For consistency, I have
followed the approach of \citet{songaila_omegaz}, summing the $\nciv$
values of individual components in each \civ complex to obtain a
single, total column density for the system.  For example, the lines
spanning $z=5.82374-5.83012$ were treated as a single system with
$\nciv=10^{14.42}$, corresponding to the sum of the individual
components listed in Table \ref{tab:sdss1030}.

The resulting distributions are shown in Figure \ref{fig:cddf} and
Table \ref{tab:cddf}.  The solid line shows the fit to the \civ CDDF
derived from the $z\sim 3$ sample of \citet{songaila_omegaz}, after
rescaling to $\Omega_M=0.3$.  The open circles indicate the $z\sim 6$
CDDF values obtained before correcting for sample incompleteness (as
described in Section \ref{sec:completeness}).  Completeness corrected
points are shown with solid dots; the corrections were made by
rescaling $\Delta X$ by the recovery fractions shown in Figure
\ref{fig:completeness}.  The horizontal error bars indicate bin sizes,
and the vertical bars indicate Poisson counting errors.

Despite the modest size of the $z\sim 6$ sample, the data do appear to
be consistent with a non-evolving form of the CDDF at $z=5-6$.  In the
primary sample with highest data quality at $z=5.72-6.21$, the data
fall directly on top of the low redshift fit.  In the lower redshift
bin the SNR is also lower so the sample is substantially incomplete
(e.g. there is only one system in the the low column density bin, and
we expect sample contamination).  However, within substantial errors
even this sample is consistent with a non-evolving distribution.

It would be stretching the data somewhat to consider these
calculations a detailed measurement of the CDDF at $z\sim 6$; a full
and accurate measurement would require both more sightlines to build
up $\Delta X$ at $\nciv \gtrsim 10^{14}$ \pcmsq ~(where our sample is
mostly complete), and higher SNR data to improve the sample
completeness for weaker lines at $\nciv \lesssim 10^{13.5}$ \pcmsq.
However, the fact that we detect any lines at all argues against a
strong downward trend in the \civ abundance between redshift 5 and 6,
and the similar overall normalization (about one system per unit
absorption path, per logarithmic unit column density) suggests that
these is no statistically significant evidence for evolution.

It is notable that within the $z=5.72-6.21$ bin, most of the \civ
absorption is concentrated at the bluest end; in fact in Section
\ref{sec:falsepos} it was suggested that for $z\gtrsim 5.9$ the data
may be statistically consistent with no \civ signal.  It is tempting
to interpret this as a hint of decline in the \civ abundance, but
there are dangers in this sort of {\em a posteriori} definition of a
redshift boundary, particularly for small sample sizes (recall that
the redshift bins for the main sample were bounded only by random
factors: a change in the noise properties of the data, and the QSO
emission redshifts).

Nevertheless, for argument's sake one can ask whether the lack of \civ
at $z\gtrsim 5.9$ should be surprising given the amount of absorption
path covered in the SDSS1030+0526 spectrum.  In the rightmost panel of
Figure \ref{fig:cddf}, we separate out the $z=5.9-6.28$ redshift
range, and indicate the upper limits derived from assuming that there
is $<1$ system at $\log(\nciv)=13.75-14.50$ (we detected none) and
$\le 4$ systems at $\log(\nciv)=13.0-13.75$ (we detected 4 systems,
but also 3 false positives).  
Taking just the strong systems at $\nciv=10^{13.75-14.5}$ which should
be easily detected, the completeness-corrected absorption pathlength
in the SDSS1030 sightline is only marginally large enough ($\Delta X =
1.16$) to expect a single strong \civ detection, so the absence of
strong \civ could be the result of a statistical fluctuation.  The
probability of observing zero strong systems in SDSS1030 at $z\gtrsim
5.9$ can be estimated from $f(\nciv)$ (the \civ CDDF) as:
\begin{equation}
P(0) = \exp\left[{{-f(\nciv) \Delta X \Delta \nciv}}\right] \approx 50\%.
\end{equation}
So, for the present, the possibility of a decline in \civ abundances
at $z\gtrsim 5.9$ remains intruiging but the sample is too small to
make conclusive statements.  If the CDDF remains constant at $z\gtrsim
6$, then there is a $\sim 75\%$ chance that a strong system will be
detected in the next sightline to be searched at $z\ge 6$.

\subsection{{\rm $\ociv$}: Contribution to Closure Density}\label{sec:omega_civ}

\begin{figure}
\epsscale{1.2}
\plotone{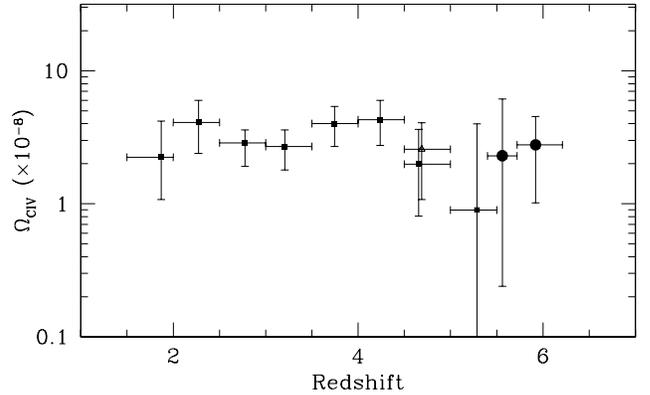}
\caption{Evolution of the \civ contribution to closure density.  New
points are shown with round dots at $z=5.40-5.72$ and $z=5.72-6.21$.
Verical error bars represent 90\% confidence intervals.  Results at
$z=1.8-5.5$ from \citet{songaila_omegaz} are shown with squares, and
the $z=4.5-5.0$ point from \citet{pettini_z5_civ} is shown with an
open triangle.  All values are scaled to $h=0.71$, $\Omega_M=0.3$, and
$\Omega_\Lambda=0.7$.}
\label{fig:omega}
\end{figure}

The integral of the CDDF (weighted by $N_\mciv$) can be used to
estimate the volume-averaged density of \civ atoms over a specified
redshift interval.  This quantity is customarily normalized by the
critical density $\rho_c$ to yield $\ociv$, which is calculated from
observed quantities as:
\begin{equation}
\ociv = {{1}\over{\rho_c}}m_\mciv {{\sum{N_\mciv}}\over{{{c}\over{H_0}}\sum{\Delta X}}}
\label{eqn:omega}
\end{equation}
where $m_\mciv$ is the mass of the \civ ion.  This integrated \civ
density provides a conveient metric for studying the redshift
evolution of \civ abundances.  \citet{songaila_omegaz} has compiled
measurements of $\ociv$ in the $z=2-5$ interval;
\citet{pettini_z5_civ} have also published a measurement at
$z=4.5-5.0$.  Figure \ref{fig:omega} displays these prior results, for
$H_0=71$ \kms Mpc$^{-1}$, and $\Delta X$ rescaled as above for
$\Omega_M=0.3$ and $\Omega_\Lambda=0.7$.

The rightmost points show the two new estimates of $\ociv$ at
$z=5.40-5.72$ and $z=5.72=6.21$ from this analysis.  Horizontal error
bars indicate the bin ranges, and vertical error bars represent 90\%
confidence intervals calculated as in \citet{pettini_z5_civ}, using
the error estimator of \citet{storrie_lombardi_omegaerr}:
\begin{equation}
{{\Delta(\ociv)}\over{\ociv}}(90\%) = 1.64\times \sqrt{{{\sum N_\mciv^2}\over{\left({\sum
N_\mciv}\right)^2}}} 
\end{equation}
The new points in Figure \ref{fig:omega} have been corrected for
sample incompleteness by scaling each line's column density in the sum
from Equation \ref{eqn:omega} by the pathlength over which it could be
detected.  It is also corrected for spurious contamination by
subtracting off the total column density of false-positive lines
identified in each bin (see Section \ref{sec:falsepos}).  These two
corrections are both of order $\sim 0.2$ dex but act in opposite
directions.

The $z\sim 6$ points follow the trend of a constant $\ociv$
established at lower redshift, and do not show evidence for a decline
at $z\gtrsim 5$.  While some hints of a declining $\ociv$ were present
in the optical data at $z\sim 4.5-5.5$, this may be an artifact of
shot noise, as was pointed out by Songaila in the work where it was
reported.  The $z=5-5.5$ range is still particularly difficult for
such measurements because of the low sensitivities of both CCDs and IR
focal plane arrays at $\lambda\sim 0.98 \mu$m, and the substantial
atmospheric foregrounds.  Both Songaila's measurement at $z=5-5.5$ and
the new point at $z=5.4-5.72$ are limited by shot noise.  A
pathlength-weighted average of the two points yields
$\ociv(5.00-5.72)=3.6^{+2.9}_{-3.5}\times 10^{-8}$, consitent with the
full redshift range, though still with substantial error bars.  In the
present sample the $z=5.72-6.21$ point is more reliable, since it
incorporates more absorption path, it is situated in a cleaner portion
of the $Y$ band, and it represents data with a higher SNR.  The error
bars for both $z\sim 6$ points are larger than for the lower redshift
data because we have only sampled two sightlines; however since
redshift path is a comoving quantity proportional to $(1+z)dz$ the
effective path per sightline increases at higher reshift for $\ociv$
measurements.

Given the resolution and SNR of the $z\sim 6$ spectra, even a single
\civ detection is enough to bring $\ociv$ up to the $\sim 10^{-8}$
level.  If the \civ density was in rapid decline at $z\gtrsim 5$, no
\civ lines should have been detected in the GNIRS spectra, and in this
case our sensitivity would have been sufficient to set upper limits
roughly a factor of 3 below the $\sim 3\times10^{-8}$ level seen at
lower redshift.  Since several \civ lines are detected, even if a
fraction of these are false-positives, it seems unlikely that they
could mask an order-of-magnitude decline in $\ociv$ at $z\sim 6$.
 
\section{Discussion}\label{sec:discussion}

\subsection{Implications for Cosmic Abundances}

The $z\sim 6$ \civ systems detected in SDSS1030+0524 and SDSS1306+0256
yield an $\ociv$ value of $\approx 3\times 10^{-8}$, which is
essentially identical to the value observed at lower redshift.
However, this measurement reflects only the \civ abundance, and does
not take into account the fraction of carbon atoms in other ionization
states.

A correction for ionization requires estimates of
$f_\mciv=n_\mciv/n_{\rm c}$---a factor which varies with redshift and
environment and can be large.  In ionization equilibrium, the exact
value of $f_\mciv$ depends upon the density ratio of ionizing photons
to baryons as well as the shape of the background radiation spectrum.
At high redshift, the ionizing background becomes less intense because
of the declining QSO luminosity function, and softer because of an
increasing contribution from galaxy light.  Also, the average gas
density increases as $(1+z)^3$, leading to a decline in the overall
ionization level.

\begin{figure}
\vskip 0.1in
\plotone{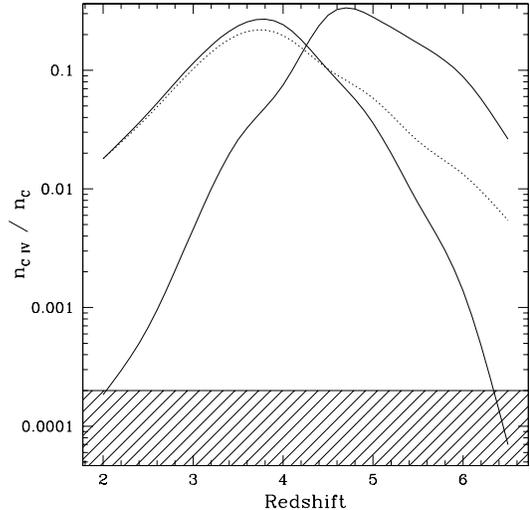}
\caption{The fraction of carbon atoms in the \civ ionization state, as
a function of redshift.  Solid lines indicate the evolution for
baryonic overdensities of $\oden=1$ and $\oden=10$, with the
$\oden=10$ curve peaking at $z\sim3.8$ and the $\oden=1$ curve peaking
at $z=4.75$.  Dotted line indicates model for $\oden=10$, but with
additional local radiation from a galaxy at a distance of $100$ kpc.
The hatched region is excluded on the basis that it would lead to
super-solar IGM abundances.}
\label{fig:ionization}
\end{figure}

The net effect of these changes is illustrated in Figure
\ref{fig:ionization}, which shows the evolution of $f_\mciv$ from
$z\sim 2-6$.  These calculations were performed using the CLOUDY
photoionization code \citep{cloudy}, with a background radiation
spectrum containing both QSOs ($f_\nu\propto \nu^{-1.8}$) and galaxies
(with 10\% escape fraction of ionizing photons) after propegation
through the IGM \citep{haardt_cuba}.  The spectra were normalized to
be consistent with measurements of the proximity effect in QSOs
\citep{scott_uv_bkgd} and the distribution of flux transmission in the
\lya forest \citep{mcdonald_uv_bkgd}.  The normalization becomes quite
uncertain approaching $z\sim 6$; in this range I have estimated its
value at by extrapolating from lower redshift data.  The two solid
lines indicate the \civ ionization fractions for gas with $\oden=1$
and $10$, where $\bar{\rho}=\Omega_b\rho_c(1+z)^3$.  At $z\sim 2$, the
intense radiation from QSOs ionizes most carbon to \cv and other
states higher than C~{\sc iv}.  As one progresses through $z\sim 4$,
the \civ fraction reaches its maximum and then declines rapidly at
$z\gtrsim 4.5$ as the carbon becomes more neutral.

The total variation in $f_\mciv$ spans 2-3 orders of magnitude from
$z\sim 2-6$ for a given overdensity.  However, it approaches a maximum
value of $f_\mciv\approx 0.3$ independent of density, though the
corresponding redshift does vary with $\oden$.  As pointed out by
\citet{songaila_omegaz}, this fact can be used to set a lower limit on
the overall carbon abundance at high redshift of $\Omega_{\rm C}
\gtrsim \ociv/0.3\approx 1\times 10^{-7}$.  This can be compared to
the overall baryon density $\Omega_b h^2=0.022$
\citep{omeara_deuterium} to obtain an volume-averaged carbon
abundance:
\begin{eqnarray}
\label{eqn:metallicity}
\left[{{\rm C}\over{\rm H}}\right] & \gtrsim & \log\left({{\Omega_{\rm C}}\over{\Omega_b}}\right)-\log\left({{m_C}\over{\mu m_H}}\right) -\log\left({{\rm C}\over{\rm H}}\right)_\odot \\
& \approx & -3.16. \nonumber
\end{eqnarray}
Here, $m_C$ and $m_H$ represent the atomic mass of carbon and
hydrogen, $\mu$ is the mean atomic mass in the IGM (assumed to be
1.3), and the last term represents the solar carbon abundance by
number, assumed to be $3.3\times 10^{-4}$
\citep{grevesse_solar_abund}.

It should be heavily emphasized that this value does not represent a
ubiquitous metallicity floor in the IGM.  Rather it is a volume
average over the entire redshift pathlength, which contains small
regions with strong \civ absorption, and large regions with no
detectable \civ.  The expectation value of $n_\mciv$ will be
substantially lower than $\Omega_\mciv \rho_c(1+z)^3$ for most
randomly chosen volumes, except for a few regions where it is much
higher than this mean value.  However, Equation \ref{eqn:metallicity}
does provide a global average which may be appropriate for comparing
against the integrated metal content, for example, within an entire
cosmological simulation volume.

As a final note, it would also be interesting to measure the abundance
of other carbon and/or silicon ions (e.g \siiv 1400, or \cii1334), since
the ratios of these ions probe both relative heavy element abundances
and the shape of the ionizing background spectrum (which is of
particular interest during reionization at $z\gtrsim 6$).
Unfortunately the SNR of the present data are not sufficient to
significantly constrain Si~{\sc iv}/\civ or other ion ratios, though
this will be an important goal of future observations.

\subsection{What does {\rm $\ociv$} Represent?}


The distribution of \civ shows several remarkable properties that
might not be expected {\em a priori}.  The most obvious of these is
the near-constant value of $\ociv$ from $z=2$ to $z=6$.  The
measurements are identical to within errors---even accounting for the
90\% confidence intervals shown in Figure \ref{fig:omega}, $\ociv$ can
vary by no more than a factor of $\sim 3$ over this complete redshift
range.  This comes despite the fact that $f_\mciv$ can vary by up to 3
orders of magnitude, and the expectation that cosmic abundances should
decline on average towards higher $z$.  In order to keep $\ociv$
constant, the evolution in ionization would need to precisely cancel
an opposite evolution in the carbon abundance, a scenario that is
possible, but which requires fine-tuning.

Second, the \civ mass is distributed very unevenly.  In the $z\sim
5.72-6.21$ bin, $43\%$ of the \civ atoms are contained within a single
system (at $z=5.823-5.830$ in SDSS1030).  A similar segregation of
\civ mass is seen at all redshifts.  This is related to the fact that
the CDDF slope is shallower than $N_\mciv^{-2}$; so that the integral
$\ociv\propto\int N_\mciv f(N_\mciv)dN_\mciv$ is dominated by the
strongest system in any sample.



At lower redshifts ($z\sim 2.5-3.5$), there is evidence that the
strongest \civ absorption systems arise in the outer halos of star
forming galaxies \citep{simcoe_feedback, kurt_winds}.  They are often
accompanied by heavy element absorption from many other ions including
\ovi, and their physical characteristics suggest a relation to
energetic feedback from galaxy formation.  Absorption from
feedback-related systems comprises at least half of the $\ociv$
integral at $z\sim 3$ \citep{simcoe_feedback, songaila_outflows}, and
these lower redshift systems share some features with the $z\sim 6$
\civ systems.  In particular the range of \civ column densities and
complexity of velocity structure seem to be similar for the strongest
\civ systems at all epochs.

This may provide one natural explanation for the constant value
of $\ociv$ with redshift.  
%
If the \civ traced by the strong systems is produced locally by star
forming galaxies, they can be enriched to $0.01-0.1Z_\odot$ levels on
short timescales.  In this case, $\ociv$ would be related more closely
to the instantaneous metal production (i.e. star formation) rate at a
given redshift, rather than the cumulative stellar mass formed prior
to the epoch of observation.  Much like $\ociv$, the star formation
rate density appears to evolve fairly little between $z=2-6$
\citep[e.g.][]{goods_lum_density,steidel_lum_density,hudf_lum_density}.
Furthermore, if the \civ absorbers reside near forming galaxies then
the spectrum of ionizing radiation that they see would be augmented by
local starlight, providing a buffer against large variations in
$f_\mciv$.

In Figure \ref{fig:ionization}, I illustrate this effect in the dotted
curve, which shows the evolution of $f_\mciv$ for the same Haardt \&
Madau diffuse background spectrum, but with an additional star forming
galaxy located $100$ kpc from the absorption system.  The galaxy
template spectrum is taken from the Starburst99 archive
\citep{starburst99}.  It has a stellar age of 300 Myr, and has been
reddened using a \citet{calzetti} extinction law with $E(B-V)=0.155$
\citep{alice_lbg_nirc}.  The total stellar mass is $10^{8}M_\odot$.

The galaxy has little effect on the ionization balance at $z\lesssim
4.5$, but at higher redshift the added galaxy flux does begin to
contribute such that $f_\mciv$ drops by only an order of magnitude
from its peak, rather than $2-3$ orders of magnitude as it would
otherwise have done.  It seems plausible that a decline in $f_\mciv$
of a factor of $\lesssim 10$ could be coupled with a factor of
$\lesssim 10$ decline in the carbon abundance at high redshift,
resulting in weak observed evolution in $\ociv$.

The other possible scenario invokes a balance of several effects to
produce a flat $\ociv(z)$, and does not require a local origin for the
observed \civ absorbers.  In this interpretation, the \civ systems
trace intergalactic regions of increasingly smaller overdensity at
higher redshift, in essence picking out the density where the \civ
ionization fraction (or $f_\mciv$ times the heavy element abundance)
peaks at any particular redshift.  One can infer from a comparison of
the solid lines in Figure \ref{fig:ionization} that this provides an
independent mechanism to suppress the variation in ionization
parameter with redshift.  It would require a fairly uniform
distribution of carbon abundance across all densities of gas in the
IGM, which although more difficult to achieve at high redshift, may
still be possible.

This explanation is difficult to test observationally, since at high
redshift one also loses information about the local density field
(because of saturation and blending within the \hi Lyman alpha
forest).  Naively one would expect $d\mathcal{N}/dX$ for \civ to
increase with redshift in this scenario (an effect which is not
observed), unless the actual chemical abundances correlate with IGM
density in just the right fashion.

This ``balancing'' hypothesis convolves the evolution of several
competing factors (characteristic densities, ionization parameters,
and chemical enrichment), so its ultimate evaluation will be best
accomplished through ``observations'' of numerical simulations which
incorporate models of chemical feedback.  Indeed,
\citet{oppenheimer_civ} have very recently run simulations where
galactic wind prescriptions motivated by local observations seem to
strike the balance required to match the $z=2-5$ observations.  In
their simulations there is evidence that $\ociv$ should start to turn
down by $z\sim 6$ in contrast to the findings presented here.
However, the level of this discrepancy is probably not alarming when
one takes into account both our measurement uncertainties and
uncertainties in the contribution from unresolved small-scale \civ
structures in the simulations.

\section{Conclusions}\label{sec:conclusions}

This paper presents moderate resolution spectra of two $z\sim 6$ QSOs,
taken with GNIRS on Gemini South.  These spectra were systematically
searched for high redshift intergalactic \civ absorption, using
objective selection criteria.  Comprehensive testing was performed to
quantify the \civ sample incompleteness as a function of $N_\mciv$, as
well as the rate of false positive \civ identifications.  Over the
primary redshift range of $z=5.72-6.21$, the sample is largely
complete at $N_\mciv\sim 10^{14}$\pcmsq, and drops uniformly to about
$5-10\%$ at $N_\mciv\sim 10^{13}$; at lower redshift the completeness
is somewhat worse because the data have a lower SNR.  However, this
sensitivity is adequate to construct a rough estimate of the \civ
column density distribution $f(N_\mciv)$, and also to measure the
integrated density $\Omega_\mciv$ at a level two to three times below
the observed value at $z=2-5$.  The main results are as follows:
\begin{enumerate}
\item{Nine \civ absorption lines are identified between $z=5.72$ and
$z=6.21$.  Three of the these lines are contained within a single
strong complex at $z=5.827\pm0.004$ toward SDSS1030+0524.  This single
system contains $43\%$ of the \civ atoms along the total combined
redshift path.  Three additional \civ lines are detected in a lower
redshift bin ranging from $z=5.40-5.72$.  After adjusting for
absorption pathlength and incompleteness, the density of lines is
similar in both redshift bins.}
\item{The spectral region at $z\ge 6$ in SDSS1030+0524 is
statistically consistent with a complete lack of \civ absorption.
However, with only one sightline sampled at present this could easily
be the result of a statistical fluctuation.}
\item{A coarse-binned representation of the $z\sim 6$ \civ column
density distribution is consistent with the redshift-invariant form
reported over the $z=2-5$ range by \citet{songaila_omegaz}.}
\item{An extension of the $\Omega_\mciv(z)$ plot to $z\sim 6$ does not
show evidence of a downturn in the integrated \civ abundance, at least
within the factor of $\sim 3$ allowed by measurement errors.}
\end{enumerate}

Over such a large redshift range, it is remarkable to derive a
constant form for the \civ column density distribution and also for
$\ociv$.  The baryonic density, ionizing background spectrum, and
metallicity should all evolve substantially between $z=2-6$, so it
would represent an interesting coincidence if these effects all
cancelled.  In principle it is possible to mask the effect of
evolution in the \civ ionization fraction if the \civ absorbers are
associated with regions of lower overdensity at higher redshifts.  In
this case one might anticipate an increase in $d\mathcal{N}/dX$ at
higher z, but this is not observed.  Because of the complex interplay
of many evolving factors, a full evaluation of this explanation for
the constancy of $\ociv$ with $z$ requires comparison with numerical
simulations \citep[e.g.][]{oppenheimer_civ}.

An attractive alternative explanation is that the $\ociv$ measurements
are dominated by strong absorbers whose properties are affected by
local chemical and radiative feedback from nearby galaxies.  This
association appears to be present at $z\sim 2.5$, where luminous
galaxies are seen directly in the neighborhood of many/most strong
\civ systems.  Although the same galaxies cannot be seen at high
redshift, they would in principle provide a mechanism to rapidly
enrich the \civ absorbing regions, and their radiation would also
regulate the ionization balance of the carbon, preventing it from
becoming too neutral for detection in \civ.

Local feedback models become more compelling at the highest
redshifts, where the elapsed time betewen the onset of reionization
($z\sim 11$) and the epoch of observation becomes quite short.  Over
timescales of $400-500$ Myr, there would be just enough time to form
an early galaxy population, build up its stellar mass, and eject heavy
elements from its earliest supernovae into a region $\lesssim 100$ kpc
in size.  In this scenario there could be large metal-free regions of
the IGM, but \civ observations at present sensitivities would only probe
the regions where local production was important.  In time, IR spectra
with higher resolution and SNR should be able to track the evolution
of weaker \civ lines in the tenuous IGM, and these systems will
provide a more sensitive test of local versus gloal enrichment
scenarios in the early universe.


\acknowledgements

I would like to extend special thanks to the Gemini Observatory staff,
and particularly Dick Joyce, for their assistance in planning and
executing the observations presented here.  Thanks as well to
Hsiao-Wen Chen for providing the code used for the curve-of-growth
analysis, and to George Becker and Wal Sargent for allowing use of
their optical spectrum of SDSS1030+0524.  Thanks to Scott Burles and
John O'Meara for advice on the data reduction and interpretation, and
to Romeel Dav{\'e} and Ben Oppenheimer for interesting discussions about
this work's relationship to their numerical simulations.  The data for
this program were obtained through Gemini allocations GS-2005A-Q-4 and
GS-2006A-Q-9.  The research was supported with funds from the
Pappalardo Fellowships in Physics program at MIT, and an AAS small
research grant.

\bibliography{z6civ}

\clearpage
\input{tab1}
\input{tab2}
\input{tab3}

\end{document}

%% file: tab1.tex
\begin{deluxetable}{c c c c c c c c c c}
\tablewidth{0pc}
\tablecaption{\civ Lines Detected Toward SDSS1306+0356}

\tablehead{{$z_{\rm \mciv}$} & {$\sigma_z$} & {$W_{\rm r, 1548}$} & {$\sigma_W$} & {$W_{\rm r, 1551}$} & {$\sigma_W$} & {\rm EW Ratio} & {$\sigma_{\rm ratio}$} & {$\log(\nciv)$} & {$N_{\rm comp}$}}

\startdata
5.45153 & 0.00015 & 104 & 14 & 42  & 13 & 2.48 & 0.82 & $13.15$ & 2 \\
5.67430 & 0.00011 & 155 & 16 & 134 & 21 & 1.17 & 0.22 & $14.30$ & 1 \\
5.92718 & 0.00011 & 80  & 11  & 30  & 4  & 2.72 & 0.54 & $12.90$ & 2 \\
\enddata
\label{tab:sdss1306}
\end{deluxetable}

%% file: tab2.tex
\begin{deluxetable}{c c c c c c c c c c}
\tablewidth{0pc}
\tablecaption{\civ Lines Detected Toward SDSS1030+0524}

\tablehead{{$z_{\rm \mciv}$} & {$\sigma_z$} & {$W_{\rm r, 1548}$} & {$\sigma_W$} & {$W_{\rm r, 1551}$} & {$\sigma_W$} & {\rm EW Ratio} & {$\sigma_{\rm ratio}$} & {$\log(\nciv)$} & {$N_{\rm comp}$}}

\startdata
5.45886 & 0.00015 & 181 & 29 & 117 & 19 & 1.54 & 0.35 & $13.65$ & 2 \\
5.73318 & 0.00005 & 254 & 13 & $\le 197$ & 14 & $\ge 1.29$ & 0.11 & $13.85$ & 2 \\
5.74313 & 0.00008 & $\le 164$ & 13 & 117 & 12 & $\le 1.41$ & 0.19 & $14.10$ & 1 \\
5.82374 & 0.00005 & 288 & 12 & 160 & 13 & 1.80 & 0.17 & $13.15$ & 6 \\
5.82791 & 0.00007 & 263 & 19 & 155 & 17 & 1.69 & 0.22 & $13.40$ & 4 \\
5.83013 & 0.00007 & 257 & 16 & 146 & 15 & 1.76 & 0.21 & $13.30$ & 4 \\
5.97594 & 0.00018 & 68 & 11 & 40 & 10 & 1.67 & 0.48 & $13.40$ & 1 \\
6.13479 & 0.00017 & 65  & 9  & 26  & 9  & 2.48 & 0.97 & $13.30$ & 1 \\
6.17568 & 0.00018 & 54  & 8  & 49  & 12 & 1.09 & 0.31 & $13.30$ & 1\\
\enddata
\label{tab:sdss1030}
\end{deluxetable}

%% file: tab3.tex
\begin{deluxetable}{c c | c c c c | c c c c}
\tablewidth{0pc}
\tablecaption{\civ Column Density Distribution by Redshift}

\tablehead{ & & \multicolumn{8}{c}{$\log(N_\mciv)$} \\ 
& & \multicolumn{4}{c}{{$13.00-13.75$}} & \multicolumn{4}{c}{$13.75-14.50$} \\ 
{$\Delta z$} & {$\Delta X_{\rm full}$\tablenotemark{1}} & { $f_{\rm comp}$\tablenotemark{2}} & {$\Delta X_{\rm eff}$\tablenotemark{3}} & {$N_{\rm lines}$} & {$\log f(N_\mciv)$\tablenotemark{4}} & { $f_{\rm comp}$} & {$\Delta X_{\rm eff}$} & {$N_{\rm lines}$} & {$\log f(N_\mciv)$}} 

\tablecolumns{10} 

\startdata 

5.40-5.72 & 2.98 & 0.193,0.084 & 0.412 & 1 & -13.27 & 0.624,0.345 & 1.440 & 2 & -14.27 \\
5.72-6.21 & 3.35 & 0.500,0.318 & 1.247 & 4 & -13.16 & 0.880,0.709 & 2.546 & 3 & -14.34 \\ 
5.90-6.21 & 1.64 & 0.379,0.350 & 0.578 & 4 & -12.83 & 0.848,0.694 & 1.162 & 0 & $<-14.48$ \\ 

\enddata

{\tablenotetext{1}{\small Full absorption path (defined by Equation
\ref{eqn:pathlength}) along redshift range assuming 100\% sample
completeness.}}

\tablenotetext{2}{\small Compleness along each sightline for
given range in redshift and $N_\mciv$.  First value is for
SDSS1306+0356, second is for SDSS1030+0524.}

\tablenotetext{3}{\small Effective redshift path for bin, determined
by applying the corrections given in previous column.}

\tablenotetext{4}{\small Column density distribution function value.}
\label{tab:cddf}
\end{deluxetable}

%% file: ms.bbl
\begin{thebibliography}{33}
\expandafter\ifx\csname natexlab\endcsname\relax\def\natexlab#1{#1}\fi

\bibitem[{{Adelberger} {et~al.}(2003){Adelberger}, {Steidel}, {Shapley}, \&
  {Pettini}}]{kurt_winds}
{Adelberger}, K.~L., {Steidel}, C.~C., {Shapley}, A.~E., \& {Pettini}, M. 2003,
  \apj, 584, 45

\bibitem[{{Calzetti} {et~al.}(1994){Calzetti}, {Kinney}, \&
  {Storchi-Bergmann}}]{calzetti}
{Calzetti}, D., {Kinney}, A.~L., \& {Storchi-Bergmann}, T. 1994, \apj, 429, 582

\bibitem[{{Djorgovski} {et~al.}(2001){Djorgovski}, {Castro}, {Stern}, \&
  {Mahabal}}]{george_reionization}
{Djorgovski}, S.~G., {Castro}, S., {Stern}, D., \& {Mahabal}, A.~A. 2001,
  \apjl, 560, L5

\bibitem[{{Elias} {et~al.}(1998){Elias}, {Vukobratovich}, {Andrew}, {Cho},
  {Cuberly}, {Don}, {Gerzoff}, {Harmer}, {Harris}, {Heynssens}, {Hicks},
  {Kovacs}, {Li}, {Liang}, {Moon}, {Pearson}, {Plum}, {Roddier}, {Tvedt},
  {Wolff}, \& {Wong}}]{elias_gnirs}
{Elias}, J.~H., {Vukobratovich}, D., {Andrew}, J.~R., {Cho}, M.~K., {Cuberly},
  R.~W., {Don}, K., {Gerzoff}, A., {Harmer}, C.~F., {Harris}, D., {Heynssens},
  J.~B., {Hicks}, J., {Kovacs}, A., {Li}, C., {Liang}, M., {Moon}, I.~K.,
  {Pearson}, E.~T., {Plum}, G., {Roddier}, N.~A., {Tvedt}, J., {Wolff}, R.~J.,
  \& {Wong}, W.-Y. 1998, in Proc. SPIE Vol. 3354, p. 555-565, Infrared
  Astronomical Instrumentation, Albert M. Fowler; Ed., ed. A.~M. {Fowler},
  555--565

\bibitem[{{Fan} {et~al.}(2004){Fan}, {Hennawi}, {Richards}, {Strauss},
  {Schneider}, {Donley}, {Young}, {Annis}, {Lin}, {Lampeitl}, {Lupton}, {Gunn},
  {Knapp}, {Brandt}, {Anderson}, {Bahcall}, {Brinkmann}, {Brunner}, {Fukugita},
  {Szalay}, {Szokoly}, \& {York}}]{fan_z6qsos_paper3}
{Fan}, X., {Hennawi}, J.~F., {Richards}, G.~T., {Strauss}, M.~A., {Schneider},
  D.~P., {Donley}, J.~L., {Young}, J.~E., {Annis}, J., {Lin}, H., {Lampeitl},
  H., {Lupton}, R.~H., {Gunn}, J.~E., {Knapp}, G.~R., {Brandt}, W.~N.,
  {Anderson}, S., {Bahcall}, N.~A., {Brinkmann}, J., {Brunner}, R.~J.,
  {Fukugita}, M., {Szalay}, A.~S., {Szokoly}, G.~P., \& {York}, D.~G. 2004,
  \aj, 128, 515

\bibitem[{{Fan} {et~al.}(2001){Fan}, {Narayanan}, {Lupton}, {Strauss}, {Knapp},
  {Becker}, {White}, {Pentericci}, {Leggett}, {Haiman}, {Gunn}, {Ivezi{\'c}},
  {Schneider}, {Anderson}, {Brinkmann}, {Bahcall}, {Connolly}, {Csabai}, {Doi},
  {Fukugita}, {Geballe}, {Grebel}, {Harbeck}, {Hennessy}, {Lamb}, {Miknaitis},
  {Munn}, {Nichol}, {Okamura}, {Pier}, {Prada}, {Richards}, {Szalay}, \&
  {York}}]{fan_z6qsos_paper1}
{Fan}, X., {Narayanan}, V.~K., {Lupton}, R.~H., {Strauss}, M.~A., {Knapp},
  G.~R., {Becker}, R.~H., {White}, R.~L., {Pentericci}, L., {Leggett}, S.~K.,
  {Haiman}, Z., {Gunn}, J.~E., {Ivezi{\'c}}, {\v Z}., {Schneider}, D.~P.,
  {Anderson}, S.~F., {Brinkmann}, J., {Bahcall}, N.~A., {Connolly}, A.~J.,
  {Csabai}, I., {Doi}, M., {Fukugita}, M., {Geballe}, T., {Grebel}, E.~K.,
  {Harbeck}, D., {Hennessy}, G., {Lamb}, D.~Q., {Miknaitis}, G., {Munn}, J.~A.,
  {Nichol}, R., {Okamura}, S., {Pier}, J.~R., {Prada}, F., {Richards}, G.~T.,
  {Szalay}, A., \& {York}, D.~G. 2001, \aj, 122, 2833

\bibitem[{{Fan} {et~al.}(2002){Fan}, {Narayanan}, {Strauss}, {White}, {Becker},
  {Pentericci}, \& {Rix}}]{fan_reionization}
{Fan}, X., {Narayanan}, V.~K., {Strauss}, M.~A., {White}, R.~L., {Becker},
  R.~H., {Pentericci}, L., \& {Rix}, H. 2002, \aj, 123, 1247

\bibitem[{{Fan} {et~al.}(2006){Fan}, {Strauss}, {Richards}, {Hennawi},
  {Becker}, {White}, {Diamond-Stanic}, {Donley}, {Jiang}, {Kim}, {Vestergaard},
  {Young}, {Gunn}, {Lupton}, {Knapp}, {Schneider}, {Brandt}, {Bahcall},
  {Barentine}, {Brinkmann}, {Brewington}, {Fukugita}, {Harvanek}, {Kleinman},
  {Krzesinski}, {Long}, {Neilsen}, {Nitta}, {Snedden}, \&
  {Voges}}]{fan_z6qsos_paper4}
{Fan}, X., {Strauss}, M.~A., {Richards}, G.~T., {Hennawi}, J.~F., {Becker},
  R.~H., {White}, R.~L., {Diamond-Stanic}, A.~M., {Donley}, J.~L., {Jiang}, L.,
  {Kim}, J.~S., {Vestergaard}, M., {Young}, J.~E., {Gunn}, J.~E., {Lupton},
  R.~H., {Knapp}, G.~R., {Schneider}, D.~P., {Brandt}, W.~N., {Bahcall}, N.~A.,
  {Barentine}, J.~C., {Brinkmann}, J., {Brewington}, H.~J., {Fukugita}, M.,
  {Harvanek}, M., {Kleinman}, S.~J., {Krzesinski}, J., {Long}, D., {Neilsen},
  E.~H., {Nitta}, A., {Snedden}, S.~A., \& {Voges}, W. 2006, \aj, 131, 1203

\bibitem[{{Fan} {et~al.}(2003){Fan}, {Strauss}, {Schneider}, {Becker}, {White},
  {Haiman}, {Gregg}, {Pentericci}, {Grebel}, {Narayanan}, {Loh}, {Richards},
  {Gunn}, {Lupton}, {Knapp}, {Ivezi{\'c}}, {Brandt}, {Collinge}, {Hao},
  {Harbeck}, {Prada}, {Schaye}, {Strateva}, {Zakamska}, {Anderson},
  {Brinkmann}, {Bahcall}, {Lamb}, {Okamura}, {Szalay}, \&
  {York}}]{fan_z6qsos_paper2}
{Fan}, X., {Strauss}, M.~A., {Schneider}, D.~P., {Becker}, R.~H., {White},
  R.~L., {Haiman}, Z., {Gregg}, M., {Pentericci}, L., {Grebel}, E.~K.,
  {Narayanan}, V.~K., {Loh}, Y.-S., {Richards}, G.~T., {Gunn}, J.~E., {Lupton},
  R.~H., {Knapp}, G.~R., {Ivezi{\'c}}, {\v Z}., {Brandt}, W.~N., {Collinge},
  M., {Hao}, L., {Harbeck}, D., {Prada}, F., {Schaye}, J., {Strateva}, I.,
  {Zakamska}, N., {Anderson}, S., {Brinkmann}, J., {Bahcall}, N.~A., {Lamb},
  D.~Q., {Okamura}, S., {Szalay}, A., \& {York}, D.~G. 2003, \aj, 125, 1649

\bibitem[{{Ferland} {et~al.}(1998){Ferland}, {Korista}, {Verner}, {Ferguson},
  {Kingdon}, \& {Verner}}]{cloudy}
{Ferland}, G.~J., {Korista}, K.~T., {Verner}, D.~A., {Ferguson}, J.~W.,
  {Kingdon}, J.~B., \& {Verner}, E.~M. 1998, \pasp, 110, 761

\bibitem[{{Giavalisco} {et~al.}(2004){Giavalisco}, {Dickinson}, {Ferguson},
  {Ravindranath}, {Kretchmer}, {Moustakas}, {Madau}, {Fall}, {Gardner},
  {Livio}, {Papovich}, {Renzini}, {Spinrad}, {Stern}, \&
  {Riess}}]{goods_lum_density}
{Giavalisco}, M., {Dickinson}, M., {Ferguson}, H.~C., {Ravindranath}, S.,
  {Kretchmer}, C., {Moustakas}, L.~A., {Madau}, P., {Fall}, S.~M., {Gardner},
  J.~P., {Livio}, M., {Papovich}, C., {Renzini}, A., {Spinrad}, H., {Stern},
  D., \& {Riess}, A. 2004, \apjl, 600, L103

\bibitem[{{Grevesse} \& {Sauval}(1998)}]{grevesse_solar_abund}
{Grevesse}, N. \& {Sauval}, A.~J. 1998, Space Science Reviews, 85, 161

\bibitem[{{Haardt} \& {Madau}(2001)}]{haardt_cuba}
{Haardt}, F. \& {Madau}, P. 2001, in Clusters of Galaxies and the High Redshift
  Universe Observed in X-rays

\bibitem[{{Leitherer} {et~al.}(1999){Leitherer}, {Schaerer}, {Goldader},
  {Delgado}, {Robert}, {Kune}, {de Mello}, {Devost}, \&
  {Heckman}}]{starburst99}
{Leitherer}, C., {Schaerer}, D., {Goldader}, J.~D., {Delgado}, R.~M.~G.,
  {Robert}, C., {Kune}, D.~F., {de Mello}, D.~F., {Devost}, D., \& {Heckman},
  T.~M. 1999, \apjs, 123, 3

\bibitem[{{McDonald} \& {Miralda-Escud{\'e}}(2001)}]{mcdonald_uv_bkgd}
{McDonald}, P. \& {Miralda-Escud{\'e}}, J. 2001, \apjl, 549, L11

\bibitem[{{O'Meara} {et~al.}(2001){O'Meara}, {Tytler}, {Kirkman}, {Suzuki},
  {Prochaska}, {Lubin}, \& {Wolfe}}]{omeara_deuterium}
{O'Meara}, J.~M., {Tytler}, D., {Kirkman}, D., {Suzuki}, N., {Prochaska},
  J.~X., {Lubin}, D., \& {Wolfe}, A.~M. 2001, \apj, 552, 718

\bibitem[{{Oppenheimer} \& {Dav{\'e}}(2006)}]{oppenheimer_civ}
{Oppenheimer}, B.~D. \& {Dav{\'e}}, R. 2006, ArXiv Astrophysics e-prints

\bibitem[{{Pettini} {et~al.}(2003){Pettini}, {Madau}, {Bolte}, {Prochaska},
  {Ellison}, \& {Fan}}]{pettini_z5_civ}
{Pettini}, M., {Madau}, P., {Bolte}, M., {Prochaska}, J.~X., {Ellison}, S.~L.,
  \& {Fan}, X. 2003, \apj, 594, 695

\bibitem[{{Rauch} {et~al.}(1996){Rauch}, {Sargent}, {Womble}, \&
  {Barlow}}]{rauch_civ_kinematics}
{Rauch}, M., {Sargent}, W.~L.~W., {Womble}, D.~S., \& {Barlow}, T.~A. 1996,
  \apjl, 467, L5+

\bibitem[{{Rousselot} {et~al.}(2000){Rousselot}, {Lidman}, {Cuby}, {Moreels},
  \& {Monnet}}]{rousselot_oh}
{Rousselot}, P., {Lidman}, C., {Cuby}, J.-G., {Moreels}, G., \& {Monnet}, G.
  2000, \aap, 354, 1134

\bibitem[{{Schaye} {et~al.}(2003){Schaye}, {Aguirre}, {Kim}, {Theuns}, {Rauch},
  \& {Sargent}}]{schaye_civ_pixels}
{Schaye}, J., {Aguirre}, A., {Kim}, T., {Theuns}, T., {Rauch}, M., \&
  {Sargent}, W.~L.~W. 2003, \apj, 596, 768

\bibitem[{{Scott} {et~al.}(2000){Scott}, {Bechtold}, {Dobrzycki}, \&
  {Kulkarni}}]{scott_uv_bkgd}
{Scott}, J., {Bechtold}, J., {Dobrzycki}, A., \& {Kulkarni}, V.~P. 2000, \apjs,
  130, 67

\bibitem[{{Shapley} {et~al.}(2001){Shapley}, {Steidel}, {Adelberger},
  {Dickinson}, {Giavalisco}, \& {Pettini}}]{alice_lbg_nirc}
{Shapley}, A.~E., {Steidel}, C.~C., {Adelberger}, K.~L., {Dickinson}, M.,
  {Giavalisco}, M., \& {Pettini}, M. 2001, \apj, 562, 95

\bibitem[{{Simcoe} {et~al.}(2004){Simcoe}, {Sargent}, \& {Rauch}}]{simcoe2004}
{Simcoe}, R.~A., {Sargent}, W.~L.~W., \& {Rauch}, M. 2004, \apj, 606, 92

\bibitem[{{Simcoe} {et~al.}(2006){Simcoe}, {Sargent}, {Rauch}, \&
  {Becker}}]{simcoe_feedback}
{Simcoe}, R.~A., {Sargent}, W.~L.~W., {Rauch}, M., \& {Becker}, G. 2006, \apj,
  637, 648

\bibitem[{{Songaila}(2001)}]{songaila_omegaz}
{Songaila}, A. 2001, \apjl, 561, L153

\bibitem[{{Songaila}(2005)}]{songaila_new_civ}
---. 2005, \aj, 130, 1996

\bibitem[{{Songaila}(2006)}]{songaila_outflows}
---. 2006, \aj, 131, 24

\bibitem[{{Spergel} {et~al.}(2006){Spergel}, {Bean}, {Dore'}, {Nolta},
  {Bennett}, {Hinshaw}, {Jarosik}, {Komatsu}, {Page}, {Peiris}, {Verde},
  {Barnes}, {Halpern}, {Hill}, {Kogut}, {Limon}, {Meyer}, {Odegard}, {Tucker},
  {Weiland}, {Wollack}, \& {Wright}}]{spergel_wmap_3year}
{Spergel}, D.~N., {Bean}, R., {Dore'}, O., {Nolta}, M.~R., {Bennett}, C.~L.,
  {Hinshaw}, G., {Jarosik}, N., {Komatsu}, E., {Page}, L., {Peiris}, H.~V.,
  {Verde}, L., {Barnes}, C., {Halpern}, M., {Hill}, R.~S., {Kogut}, A.,
  {Limon}, M., {Meyer}, S.~S., {Odegard}, N., {Tucker}, G.~S., {Weiland},
  J.~L., {Wollack}, E., \& {Wright}, E.~L. 2006, ArXiv Astrophysics e-prints

\bibitem[{{Steidel} {et~al.}(1999){Steidel}, {Adelberger}, {Giavalisco},
  {Dickinson}, \& {Pettini}}]{steidel_lum_density}
{Steidel}, C.~C., {Adelberger}, K.~L., {Giavalisco}, M., {Dickinson}, M., \&
  {Pettini}, M. 1999, \apj, 519, 1

\bibitem[{{Storrie-Lombardi} {et~al.}(1996){Storrie-Lombardi}, {McMahon}, \&
  {Irwin}}]{storrie_lombardi_omegaerr}
{Storrie-Lombardi}, L.~J., {McMahon}, R.~G., \& {Irwin}, M.~J. 1996, \mnras,
  283, L79

\bibitem[{{Thompson} {et~al.}(2006){Thompson}, {Eisenstein}, {Fan},
  {Dickinson}, {Illingworth}, \& {Kennicutt}}]{hudf_lum_density}
{Thompson}, R.~I., {Eisenstein}, D., {Fan}, X., {Dickinson}, M., {Illingworth},
  G., \& {Kennicutt}, R.~C. 2006, ArXiv Astrophysics e-prints

\bibitem[{{White} {et~al.}(2003){White}, {Becker}, {Fan}, \&
  {Strauss}}]{white_reionization}
{White}, R.~L., {Becker}, R.~H., {Fan}, X., \& {Strauss}, M.~A. 2003, \aj, 126,
  1

\end{thebibliography}
